\documentclass[journal=amlccd,manuscript=letter,layout=traditional]{achemso}

 \def\acs@type@list{letter,article,suppinfo} 
\SectionsOn 

\usepackage{amsmath,amsfonts,amssymb,mathrsfs,bm}
\usepackage{tabularx}
\usepackage{mathtools}
\usepackage{color}
\usepackage{siunitx}
\usepackage{graphicx}
\usepackage[dvipsnames]{xcolor}
\usepackage{cuted}
\usepackage{ulem}
\usepackage{marginnote}
\usepackage{hyperref}
\hypersetup{
    colorlinks=true,
    linkcolor=blue,
    filecolor=magenta,      
    urlcolor=cyan,
    citecolor=black
}
\usepackage{MnSymbol,wasysym}
\usepackage{array}   

\usepackage{cancel}

\DeclareMathOperator{\tr}{\textrm{tr}}
\renewcommand{\vec}[1]{\boldsymbol{#1}}
\newcommand{\tens}[1]{\mathbf{#1}}

\newcommand{\D}{\textrm{D}}

\newcommand{\bnabla}{\vec{\nabla}}

\newcommand{\ens}[1]{\left\langle #1 \right\rangle}
\newcolumntype{C}{>{$}c<{$}} 

\newcommand\redsout{\bgroup\markoverwith{\textcolor{red}{\rule[0.5ex]{2pt}{0.4pt}}}\ULon}

\newif\ifmarkup
\ifmarkup
\newcommand{\change}[1]{{\color{blue}#1}}
\newcommand\delete[1]{{\color{magenta}\sout{#1}}}
\newcommand\edelete[1]{{\color{magenta}\cancel{#1}}}
\newcommand\eqdel[1]{#1}
\else
\newcommand{\change}[1]{{\color{black}#1}}
\newcommand\delete[1]{}
\newcommand\edelete[1]{}
\newcommand\eqdel[1]{}
\fi

\usepackage{epstopdf}
\AppendGraphicsExtensions{.tiff}

\newcommand{\Wi}{\textrm{Wi}}
\newcommand{\kB}{k_{\textrm{B}}}
\mathchardef\mhyphen="2D 

\newcommand{\pes}{1.25}
\newcommand{\pem}{1.11}
\newcommand{\pel}{0.98}
\newcommand{\pea}{1.09}
\newcommand{\pesd}{0.11}
\newcommand{\pemd}{0.08}
\newcommand{\peld}{0.16}
\newcommand{\pead}{0.08}
\newcommand{\sfs}{0.27}
\newcommand{\sfl}{0.23}
\newcommand{\sfsd}{0.02}
\newcommand{\sfld}{0.04}
\newcommand{\ff}{0.09}
\newcommand{\ffd}{0.01}
\newcommand{\PEcinf}{8.2}

\title{Entanglement kinetics in polymer melts are chemically specific} 

\author{Benjamin E. Dolata}
\email{benjamin.dolata@nist.gov}
\affiliation{Department of Physics and Institute for Soft Matter Synthesis \& Metrology, Georgetown University, 3700 O St NW, Washington DC 20007, USA}
\altaffiliation{Materials Science and Engineering Division, National Institute of Standards and Technology, Gaithersburg, MD 20899, USA}
\author{Marco A. Galvani Cunha}
\affiliation{Department of Physics \& Astronomy, University of Pennsylvania, Philadelphia, PA 19104, USA}
\author{Thomas O'Connor}
\affiliation{Department of Materials Science and Engineering, Carnegie-Mellon University, Pittsburgh, PA 15213, USA}
\author{Austin Hopkins}
\affiliation{Department of Physics, University of California Santa Barbara, Santa Barbara, CA 93106, USA}
\author{Peter D. Olmsted}
\affiliation{Department of Physics and Institute for Soft Matter Synthesis \& Metrology, Georgetown University, 3700 O St NW, Washington DC 20007, USA}
\email{pdo7@georgetown.edu}

\date{\today}

\begin{document}
\newpage
\begin{abstract}
We investigate the universality of entanglement kinetics in polymer melts.  We compare predictions of a recently developed constitutive equation for disentanglement to molecular dynamics simulations of both united-atom polyethylene and Kremer-Grest models for polymers in shear and extensional flow.  We confirm that entanglements recover on the retraction timescale, rather than the reptation timescale. We find that the convective constraint release parameter $\beta$ is  independent of molecular weight, but that it increases with the ratio of Kuhn length $b_K$ to packing length $p$ as $\beta\sim (b_K/p)^\alpha$,  with an exponent $\alpha=1.9$, which may suggest that disentanglement rate correlates with an increase in the tube diameter. These results may help shed light on which polymers are more likely to undergo shear banding.
\end{abstract}

	
\maketitle 

The classical Doi-Edwards (DE) tube theory of entangled polymer melts approximates the many-body dynamics of a polymer molecule as a single-chain constrained within an effective tube arising from entanglements with the surrounding chains.\cite{DoiEdwards_Book}  
This approximation leads to a dynamical equation for the polymer conformation (which encodes both polymer orientation and stretch) for a fixed number of entanglements.  However, molecular dynamics simulations show that shear-induced disentanglement, often called convective constraint release (CCR)\cite{maigm-03,fkmk-06,bmk-10,sek-15,sek-16,nek-19a,sek-19,galvani2022probing}, can occur for strong flows. In molecular models \cite{glmm-03,marrucci-96,im-14,im-14b,hhhr-15,dolata2023} CCR is controlled by a parameter $\beta$, which represents the degree to which non-affine retraction within a tube leads to the elimination of an entanglement. Experiments on apparent yield \cite{Tapadia2004Nonlinear-flow-}, step shear \cite{sa-02}, repeated shear after short-time relaxation \cite{rr-13} and steady shear \cite{xu-22} have been interpreted in terms of disentanglement; and simulations have shown that the number of entanglements influences the properties of polymer materials, such as weld strength.\cite{wym-89,om-93,gppgr-13,cr-20} Hence, there is a need for a physical understanding of the mechanisms that 
drive disentanglement.

The primary goal of this work is to elucidate how the CCR rate (encoded by the parameter $\beta$) depends on polymer chemistry.  We study molecular dynamics simulations of three different model polymer ``chemistries'', using our own  simulations of flexible (F-KG) and semi-flexible Kremer-Grest (SF-KG) bead spring polymers\cite{kg-90} as well as data from three studies simulating united-atom polyethylene (UA-PE) by \citeauthor{sek-15}\cite{sek-15,sek-16,nek-19a,sek-19} (Simulation details can be found in the Supplementary Information, SI). The three molecules have different monomer densities and tube extensibilities, as quantified  by the number of Kuhn segments per entanglement $N_{eK}$,  the  number of chemical monomers per entanglement $N_e^{\textrm{mon}}$, and the characteristic ratio $C_{\infty}=\ens{R^2}/[(N_\textrm{mon}-1)\ell_b^2]$, where $R$ is the end-to-end distance of a chain, $N_\textrm{mon}$ is the number of monomers in the chain, and $\ell_b$ is the length of a bond (Table~\ref{tab:parameters}).


To characterize the CCR rate we compare the simulations to a thermodynamically-consistent constitutive model for flow-induced disentanglement of a polymer melt recently developed by two of the authors\cite{dolata2023}.  This model incorporates the Ianniruberto-Marrucci (IM) disentanglement mechanism,\cite{im-14,im-14b,ianniruberto-15} whereby entanglement removal is accelerated during flow (convective constraint release, or CCR). The CCR rate is characterized by a  parameter $\beta$, which is roughly inversely proportional to the number of retraction events required to remove an entanglement, so that a smaller $\beta$ requires more chain retraction events to effect entanglement removal. 

Entanglements are inferred from the average number of `kinks' $Z_k$  per chain,  determined using the Z1 code \cite{kroger-05}, which implements a geometric algorithm to reduce the polymer chains to their primitive paths and identify kinks as the points where multiple primitive paths touch. The number of kinks $Z_k$ is typically around twice the number of entanglements $Z$ determined from the rheological tube.\cite{maigm-03,tt-06,fkmk-06,bmk-10} This discrepancy arises from spatial correlations among kinks, and is quantified by $\zeta_Z=Z_k/Z$, which may be understood as analogous to a characteristic ratio for the tube itself.  We assume $\zeta_Z\simeq2$ th\change{r}oughout this work\change{, so that $Z=Z_k/2$ is the number of `rheological' entanglements}.

The constitutive model couples the dynamics of the conformation tensor of the tube $\tens{A}$ (an average of the second moment of the tube segment vectors\cite{dolata2023}), which encodes the deformation of the melt, to the entanglement ratio $\nu=\ens{Z_k}/Z_{k,eq}$, which is the ratio of the current number of entanglements $Z$ to the equilibrium value $Z_{eq}$.  The conformational dynamics are obtained by analogy with the Rolie-Poly model\cite{lg-03} with an additional Giesekus term to describe a finite second normal-stress difference, parameterized here by $\alpha$:\cite{dolata2023}
\begin{subequations}\label{eq:governing-equations}
	\begin{align}
	\begin{split}
	\frac{\D \tens{A} }{\D t} - \tens{A}\cdot\bnabla\vec{v} - (\bnabla\vec{v})^\intercal\cdot\tens{A} & = 
	- \frac{1}{\tau_d(\lambda)}(\tens{I} + \alpha(\tens{A}-\tens{I}))\cdot(h(\lambda)\tens{A} - \tens{I}) 
	- \frac{2}{\tau_R} \frac{h(\lambda) \lambda^2 - 1}{\lambda^2 + \lambda}\tens{A} \\
	& \quad - \frac{\zeta_Z\beta\nu}{3\lambda^2}
	\left(\frac{\tens{I}}{\tau_d(\lambda)} + \alpha\frac{\tens{A} - \tens{I}}{\tau_d(\lambda)} 
	+ \frac{2\lambda \tens{I}}{\tau_R(\lambda + 1)}\right)\cdot\tens{A}\ln\nu, 
	\end{split}\label{eq:governing-equation-A}\\
	\begin{split}
	\frac{\D \nu}{\D t}  & = -\beta\nu\left( \tens{S}:\bnabla \vec{v} - \frac{1}{\lambda}\frac{\D\lambda}{\D t}\right) 	- \frac{\ln\nu}{\tau_R},		
	\end{split}\label{eq:governing-equation-nu} \\
	\vec{\sigma} & = G_0\left(h(\lambda)\tens{A} - \tens{I}\right).
	\end{align}
\end{subequations}
Here we approximate the orientation tensor of the tube as $\tens{S}=\tens{A}/\tr\tens{A}$ and the polymer stretch as $\lambda=\sqrt{\tr\tens{A}/3}$;  $\D/\D t +\vec{v}\cdot\bnabla$ is the material time derivative, $\vec{v}$ is the velocity field of the melt,  $\vec{\sigma}$ is the stress tensor, $\tau_R$ and $\tau_{d,eq}$ are respectively the equilibrium Rouse and equilibrium reptation times, and $\tau_{d,eq}$ is computed from the Likhtman relation\cite{lm-02} (see \citet{dolata2023} and the SI for details).  The non-equilibrium reptation time
\begin{equation}
	\frac{1}{\tau_d(\lambda)} = \frac{1}{\tau_{d,eq}} + \frac{Z_{eq}}{Z_{eq}\nu+1}\frac{\beta\nu}{\tau_R\lambda}\frac{h(\lambda)\lambda^2-1}{\lambda + 1}
\end{equation}
arises self-consistently from the rate of entanglement removal at the chain ends, and accounts for thermal constraint release.\cite{lg-03,dolata2023}  The plateau modulus $G_0=Z_{eq}n\kB T$ can be determined from the equilibrium number of entanglements (here $n$ is the number density of polymer chains).  The dimensionless FENE (finitely extensible nonlinear elastic)-spring constant is \cite{cohen-91,sbm-09}  
\begin{equation}
    h(\lambda) = 1 + \frac{2}{3}\frac{\lambda^2 - 1}{\lambda_\textrm{max}^2 - \lambda^2},
\end{equation}
where the maximum stretch is $\lambda_\textrm{max} = \sqrt{N_{eK}}$.  The Giesekus parameter $\alpha$ ensures a non-zero second normal-stress difference.

The CCR rate $\beta$ is the only fitting parameter; $Z_{k,eq}$, $\tau_R$, $\tau_{d,eq}$, and $\lambda_\textrm{max}$ can be measured or computed from linear response behavior, and $\alpha$ can be measured from weakly non-linear rheology. In contrast to the assumptions of previous models 
\cite{im-14,im-14b,ianniruberto-15,hhhr-15,mbp-15,mmp-18} , 
\citet{dolata2023} showed the entanglement ratio relaxes according to the  
Rouse time, rather than the reptation time. This is consistent with molecular dynamics simulations\cite{ohr-19}. 

The CCR parameter $\beta$ of Eq.~\eqref{eq:governing-equation-nu}  was introduced by \citet{marrucci-96} to quantify the acceleration of the reptation rate due to the removal of entanglements in response to non-affine motion of the primitive path (distinct from the chain itself). It was then incorporated by \citet{im-14} into the dynamics for $\nu$.  In the GLaMM model CCR is determined self-consistently by equating the rate of removal of entanglements at the chain ends due to retraction with the rate of tube renewal in the tube interior \cite{glmm-03}, under the explicit assumptions that the tube persistence length is unchanged in flow and thus the number of entanglements is always proportional to the primitive path length, and that entanglements are two body interactions. The CCR dynamics postulated by \citet{im-14} relaxes this requirement. \citet{hhhr-15} introduced the concept of `entanglement stripping' to describe the loss of entanglements due to relative motion of polymer chains caused by flow and chain retraction, in addition to the steady-state renewal and removal of entanglements incorporated in the GLaMM model.

We can estimate the dependence of $\beta$ on polymer chemistry using physical arguments.The CCR parameter controls the rate of the release of topological constraints between neighboring chains.  This suggests that $\beta$ depends on Kuhn length $b_K,$ packing length $p$, and $\ens{R^2}$, the three material properties that capture the packing geometry of \change{simple monodisperse} \delete{all} melts.\cite{qin-24,hk-20,milner-20}  Furthermore, $\beta$ should be (approximately) independent of molecular weight (and hence $\ens{R^2}$), because the Ianniruberto-Marrucci mechanism is independent of  chain length. This suggests a functional form
\begin{equation}
     \beta = \beta\left(\frac{b_K}{p}\right). \label{eq:beta_b_p}
\end{equation}
Increasing $b_K/p$ will increase the ratio of the tube diameter $a$ to the packing length \cite{dkh-22}, which suggests that $\beta(b_K/p)$ is an increasing function of its argument. Physically, disentanglement is easier when the confining tube is large compared to the characteristic separation between monomers.

We test the hypothesized form of Eq.~\eqref{eq:beta_b_p} by estimating  $\beta$ from molecular dynamics simulations.  There are two ways to use the solution of Eq.~\eqref{eq:governing-equation-nu}  to obtain  $\beta$  from simulated data of the steady state entanglement ratio  $\nu({\Wi_R})$:
\begin{itemize}
\item[(A)] A full calculation of Eqs.~(\ref{eq:governing-equations}) (where $\tens{S}=\tens{S}^A\equiv\tens{A}/\tr\tens{A}$) for the stress and entanglement dynamics, which requires a constitutive equation for the conformational tensor, can be compared with the simulated stress and entanglement number. In this method one should use model polymer parameters determined from  simulations in the linear regime, aside from the anisotropy parameter $\alpha$ which is determined by the measured (simulated) normal stresses. This calculation simultaneously tests validity of stress predictions as well as disentanglement, and thus will fail if the constitutive relation is inaccurate.
\item[(B)] Alternatively, 
the equation of motion for the entanglement dynamics (Eq.~\ref{eq:governing-equation-nu}) has an exact solution in steady state 
that relates structure, flow rate, and disentanglement,\cite{dolata2023} 
\begin{equation}\label{eq:nu-steady2}
\nu_{\textrm{steady}}=
\exp\left[-W(\beta\tau_R \,\tens{S}\textbf{:}\mathbf{\nabla}\vec{v})\right],
\end{equation}
where  the Lambert $W$ function $W(x)$ obeys $We^W=x$. This can be fitted to simulation results for $\nu$ and $\tens{S}$ to find $\beta$.  This relation holds for both extensional and shear flows. 
To compute \delete{this} \change{the r.h.s. of \eqref{eq:nu-steady2}} from molecular dynamics simulations one needs to calculate the  orientation tensor.  Unfortunately, it is almost impossible to accurately calculate the conformation tensor $\tens{A}$, based on tube segments, from simulations, in order to compute $\tens{S}$.  Hence we compute the orientation tensor as
\begin{equation}\label{eq:orientation}
    \tens{S}^B \equiv \ens{\vec{u}_{N_e^\textrm{mon}}\vec{u}_{N_e^\textrm{mon}}},
\end{equation}
where $\vec{u}_{N_e^\textrm{mon}}$ is the unit vector pointing from a monomer (say $i$) to a second monomer located at a distance $N_e^\textrm{mon}$ from the first (i.e. $i\pm N_e^\textrm{mon}$). This is the length scale over which the tube decorrelates, and is approximately equivalent to two kinks,\cite{maigm-03,tt-06,fkmk-06,bmk-10} and can be easily calculated from simulations. $\tens{S}^B$, like $\tens{A}$, is a mean\change{-}field quantity, since the net tube orientation fluctuates more at the ends than in the middle.
\end{itemize}
Method A tests the full constitutive model, and the quality of the fit depends on the accuracy of the model and of the measured model inputs $(\tau_{d,eq},\tau_R,N_{eK}$, \textit{etc}.), while Method B is less restrictive and tests the accuracy of the structural predictions of Eq.~\ref{eq:governing-equation-nu} for entanglement dynamics. Hence, Method A, which is independent of the assumed constitutive model, might be expected to give a more accurate measurement of $\beta$ from simulations.

We first use Method A to estimate $\beta$ for UA-PE and KG melts. The UA-PE simulation data were obtained from recent literature\cite{sek-15,sek-16,nek-19a,sek-19}, while we performed simulations of the F-KG \change{model using 500 monomers per chain}, and SF-KG model \change{using 250 or 500 monomers per chain}. 
 We calculate the entire coupled constitutive equation set (Eq.~\ref{eq:governing-equations}) using the parameters in Table~\ref{tab:parameters}, which we then compare with the simulations. 
\begin{table*}
		\begin{tabular}{cCCCCCC|CCCCC}
	\hline\hline\\[-13truept]
	Chemistry & Z_{k,eq}  & N_{eK} & N_e^{\textrm{mon}}  & C_{\infty}^{\ast} & \dfrac{b_K^\dagger}{p} & \dfrac{p}{\ell_b} &
 \lambda_\textrm{max} & \beta_\textrm{shear} & \beta^B& \beta_\textrm{ext} & \beta_\textrm{ext}^B\\[8.0truept]\hline
	    F-KG & 12 & 40 & 86 & 1.9 & 2.8 & 1.9 &
        6.3   & \ff\pm\ffd & 0.08 
      & - & -\\
	    SF-KG-250 & 15 & 8.8 & 28 & 2.9 & 6.8 & 3.0  &
        3.0   & \sfs\pm\sfsd & -  & - & -\\
		SF-KG-500 & 31 & 8.8 & 28 & 2.9 & 6.8 & 3.0&
  3.0  &  \sfl\pm\sfld & 0.23 
  & 0.76 
  &  0.76 
  \\
		UA-PE\cite{sek-15} & 9.0 & 6.0& 65 & \PEcinf^{\ast} & 12 & 0.9 &
      2.45   &  \pes\pm\pesd & 2.0  & - & - \\
		UA-PE\cite{sek-16} & 16.4 &6.2  & 73 & \PEcinf & 12 & 0.9&
      2.49  & \pem\pm\pemd & -  & -& -  \\
		UA-PE\cite{nek-19a,sek-19} & 24.8 & 6.4 & 72 & \PEcinf & 10 & 1.0 &
    2.53  & \pel\pm\peld & 2.0  & - & -\\\hline\hline
	\end{tabular}
\caption{Parameters of the simulated polymers and fitted CCR parameters $\beta$. 
The packing length $p$ has been expressed either in units of the Kuhn step $b_K$ or the bond length $\ell_b$ between particles (KG) or united atoms (UA-PE). The parametrization is discussed in the SI; for the KG chains, $Z_{k,eq}$ was measured using the Z1 code\cite{kroger-05}, the characteristic ratio $C_\infty$ was obtained from \citet{auhl2003equilibration}, all other quantities are taken from \citeauthor{ekfhs-20}\cite{ekfhs-20} The CCR parameters $\beta_{\textrm{shear}}, \beta_{\textrm{ext}}$ are obtained using Method A, while $\beta^B$ \change{and $\beta^B_\textrm{ext}$ are} \delete{is} obtained \delete{(for shear)} using Method B. $^{\dagger}$We use the relation between the packing length $p$ and the bending modulus of KG chains given by \citeauthor{ekfhs-20}\cite{ekfhs-20} $^{\ast}$The characteristic ratio $C_{\infty}$ for UA-PE chains is obtained from the data in \citeauthor{fklk-09}\cite{fklk-09}.  We use $\alpha=0.5$ in all cases. \change{The F-KG simulations have chains of length $N_{\textrm{mon}}=500$, while the SF-KG simulations have chains of lengths $N_{\textrm{mon}}=250, 500$. Note that $Z=Z_k/2$ is the number of `rheological' entanglements}} \label{tab:parameters}
\end{table*}
\begin{figure}
	\centering
	\includegraphics[width=0.95\textwidth]{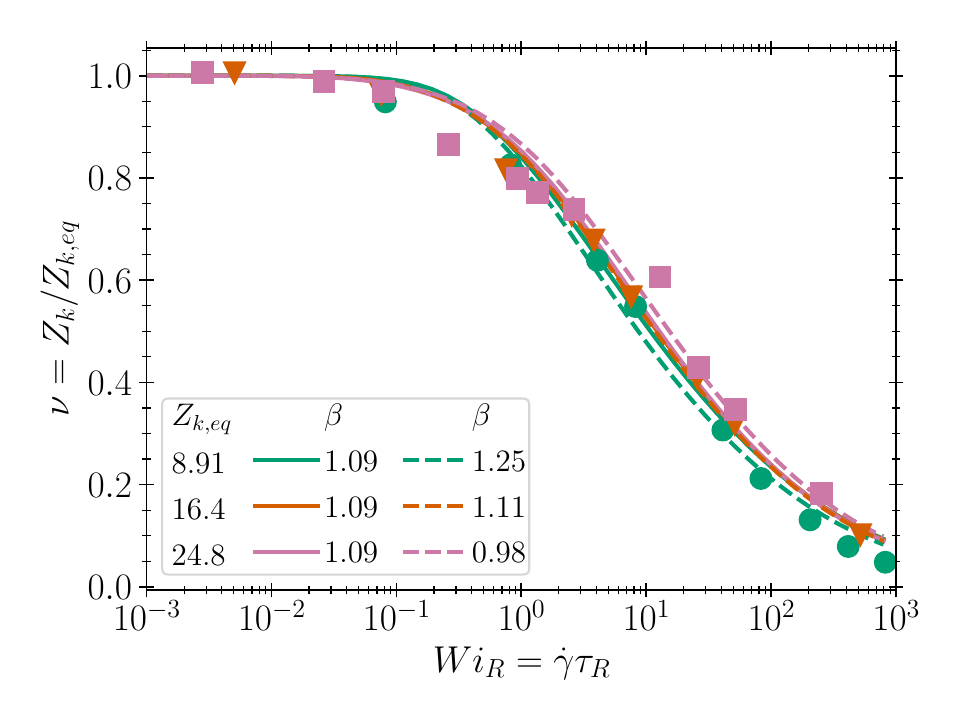}
	\caption{Comparison of our analytic model (lines, Eq.~\ref{eq:nu-steady2}) with molecular simulation data (filled \delete{circles}\change{symbols}) of disentanglement under steady state shear for three different molecular weights of UA-PE, for $Z_{k,eq}=9.0$,\cite{sek-15} $16.4$,\cite{sek-16} and $24.8$.\cite{nek-19a,sek-19}  Solid lines are fits with  the same  $\beta=\pea$, while  dashed lines are best fit values.  }
 \label{fig:khomamiZ}
\end{figure}
We first examine the influence of molecular weight on polymer disentanglement, based on the UA-PE  simulations\cite{sek-15,sek-16,nek-19a,sek-19}. Figure~\ref{fig:khomamiZ} shows the simulated degree of disentanglement in  steady shear flow as a function of $\Wi_R=\dot\gamma\tau_R$, where $\dot\gamma$ is the shear rate. We  determine $\beta$ with a least-squares regression restricted to $\Wi_{R}\geq0.1$, and find $\beta=\pes\pm\pesd,\pem\pm\pemd,$ and $\beta=\pel\pm\peld$ for $Z_{k,eq}=9.0$,\cite{sek-15} $16.4$,\cite{sek-16} and $24.9$,\cite{nek-19a,sek-19} respectively, where the  uncertainty represents 95~\% confidence.  
The small decrease in $\beta$ with increasing $Z_{k,eq}$ coincides with a slight increase in $N_{eK}$ (c.f. Table~\ref{tab:parameters}). However, all molecular weights are consistent with a single value $\beta=\pea\pm\pead$. Thus, molecular weight does not play an important role in entanglement kinetics.

\begin{figure}
	\centering
	\includegraphics[width=0.98\textwidth]{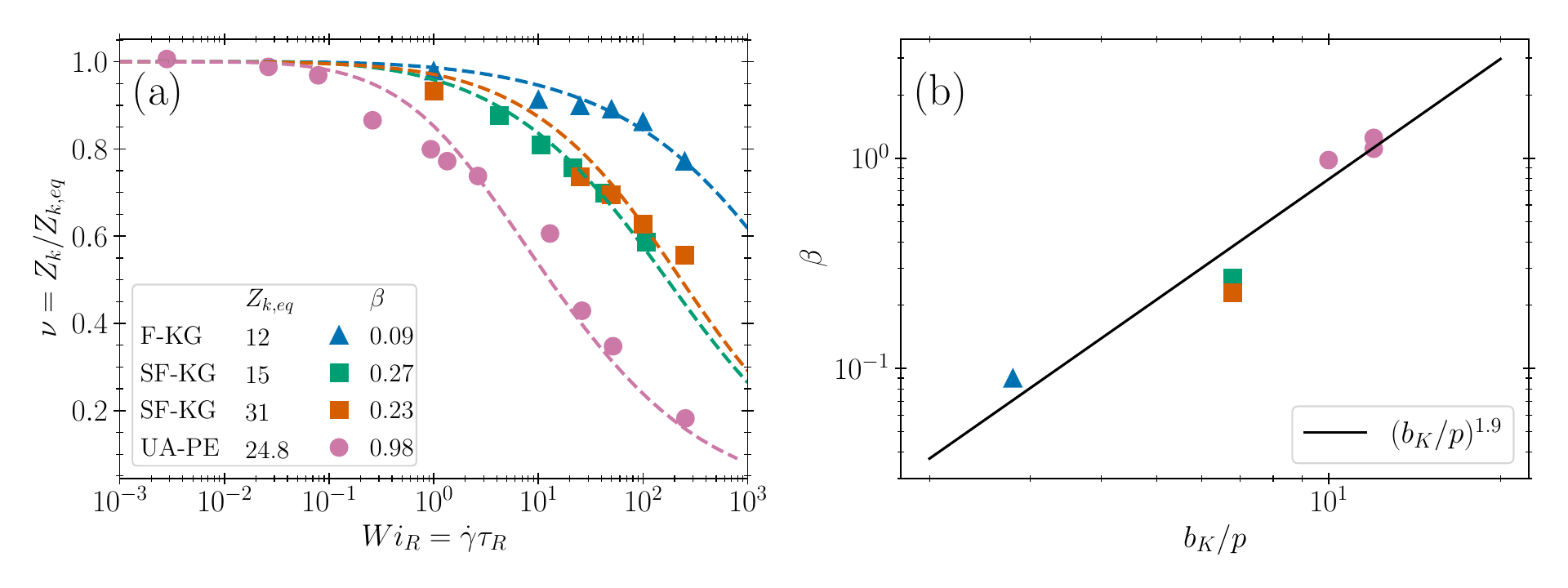}
	\caption{(a) Steady state disentanglement under shear flow for the SF-KG and UA-PE melts.  Filled symbols are molecular dynamics simulations and solid lines are model \change{fits} \delete{predictions} to determine $\beta$ \change{using Method A}. (b) CCR parameter $\beta$ as a function of $b_K/p$. Symbols represent the data from Table~\ref{tab:parameters} and the black line is a power law fit.}\label{fig:Marco-steady-state}
\end{figure}

We next compare disentanglement in steady-state shear between SF-KG and  UA-PE simulations (Figure~\ref{fig:Marco-steady-state}).  The UA-PE melt is more disentangled than the SF-KG melt at a given Weissenberg number $\Wi_R$. Consequently, UA-PE displays a best fit $\beta=\pel$ that is significantly larger than $\beta = \sfl$ for SF-KG. This implies that more retraction events are required to remove an entanglement in the SF-KG melt than in the UA-PE melt. Consistent with our hypothesized functional form  in Eq.~\eqref{eq:beta_b_p}, stiffer polymers show a larger $\beta$ (Table~\ref{tab:parameters}),  and 
the data is approximately described by a power law (Figure~\ref{fig:Marco-steady-state}(b))
\begin{equation}\label{eq:power-law}
    \beta\sim \left(\frac{b_K}{p}\right)^{1.9}.
\end{equation}
Physically, this suggests that stiffer melts will disentangle more readily than more flexible melts.  This power law may suggest that $\beta$ increases with the ratio of tube diameter to the p\change{a}cking length, which scales as $a/p\sim (b_K/p)^{1/2}$ in the semi-flexible regime \cite {milner-20,dkh-22}. We regard the observed power law as approximate, as we do not have enough data to rigorously justify the expression.  In general, we expect \change{the functional form of} $\beta(b_K/p)$ to vary between the flexible, semiflexible, and stiff regimes in a manner similar to the plateau modulus \cite{milner-20,dkh-22}


\begin{figure}
	\centering
	\includegraphics[width=0.95\textwidth]{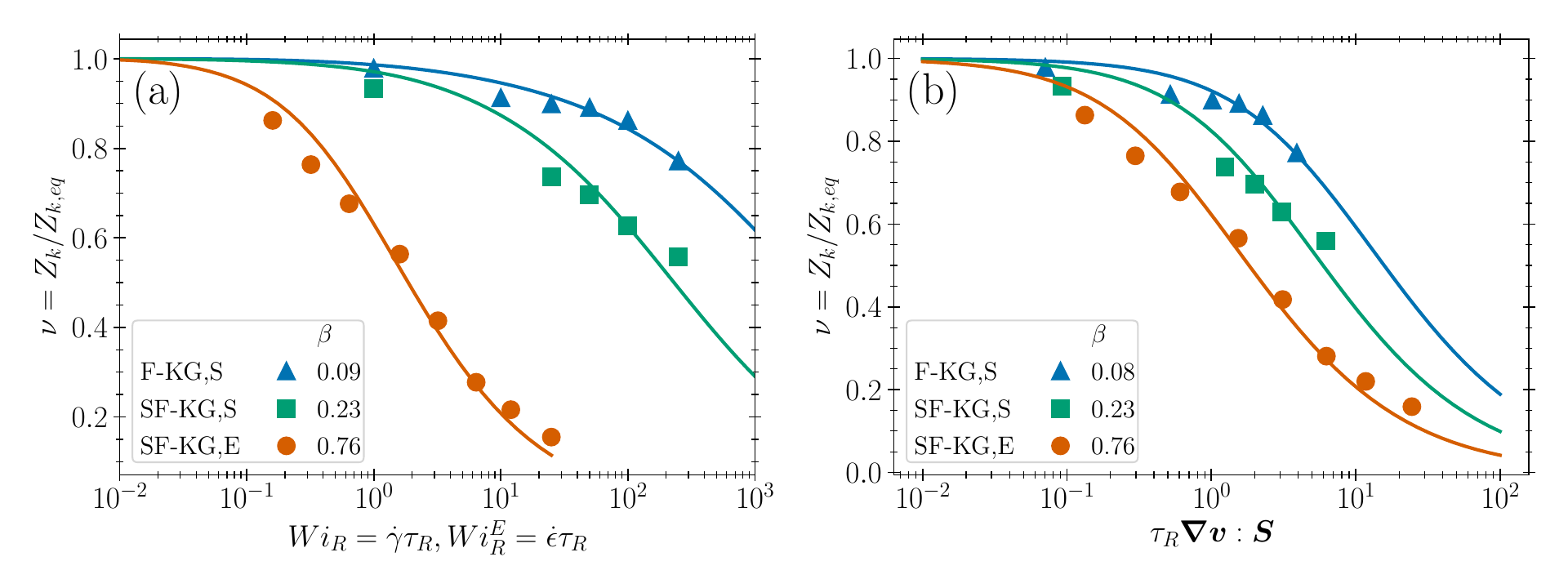}	\caption{Disentanglement of the \change{F-KG and SF-KG melts, both with 500 monomers,} \delete{F-KG 500 melt and SF-KG 500 melt} under shear (S) at constant shear rate $\dot\gamma$ and extension (E) under constant extension rate $\dot{\epsilon}$.  Filled symbols are simulation results, and solid lines are the best fit from Method A (a) and Method B (b).}\label{fig:MethodAB}
\end{figure}
We next compare Methods A and B in Figure~\ref{fig:MethodAB} using \change{the} \delete{our} F-KG \delete{500} and SF-KG melts, for disentanglement under steady-state shear (S) at rate $\dot\gamma$ and steady 3D uniaxial extension (E) at constant Hencky strain-rate $\dot\epsilon.$  Methods A and B will necessarily produce almost identical  fits in extensional flow because the molecules align nearly completely in the flow direction, so that $\tau_R\boldsymbol{\nabla v:}\tens{S}\simeq \dot{\epsilon}\tau_R$. 

The best fit value for $\beta$ is larger for extension than for shear.  The disagreement for different flow types indicates that the model \cite{dolata2023} is missing important physics, which could be due to  several potential reasons.  (1) This \change{is a} mean\change{-}field model, \change{and} assumes a spatially and dynamically homogeneous melt, whereas strong extensional flows can lead to dynamic and spatial heterogeneity,
such as separation into domains of coiled (entangled) and stretched (unentangled) states.\cite{sek-18extension}. (2) Spatial heterogeneities in the entanglement distribution could lead to inhomogeneous chain deformation in extension analogous to that seen in polymer networks \cite{mendes1991experimental}, which is not accounted for here. (3) 
Individual chains tumble in shear flow, while retraction dominates in extension, which leads to narrower distributions of conformations in extension and perhaps more accurate mean\change{-}field models.
(4) Entanglements are treated as fixed, pointlike constraints,  while in reality they fluctuate about some mean position, while under tension.  Such fluctuating tensions are expected to have a tensorial (quadrupolar) nature that can be expected to differ between shear and extension.
(5) Constraint release is assumed to be independent of the current state of the polymer conformation; however, the differences between polymer conformations in shear and extension could lead to multi-chain effects such that a given entanglement could be easier to remove in extension than in shear, which would lead to more rapid CCR and thus a larger $\beta$.  (6) Entanglements are treated in a mean\change{-field} sense, while it has been shown that the distribution of entanglements along a chain is \change{inhomogeneous} following step elongation\delete{ is inhomogeneous}.\cite{hk-18} The inhomogeneity may be expected to differ between shear and extension. Accounting for such inhomogeneity would require a multimode model, perhaps constructed in the manner of the GLaMM model.\cite{glmm-03}

We could not  apply Method B to the UA-PE data due to the absence of published data for $\tens{S}^B$.  Instead, we approximate $\tens{S}^B$ using available data\cite{dolata2023}.  For the simulations of $Z_k=9.0$\cite{bmk-10}  we  approximate $\tens{S}^B\simeq \tens{A}_{ee}/\tr\tens{A}_{ee}$, where \citet{bmk-10}  calculated $\tens{A}_{ee}=\ens{\vec{R}_{ee}\vec{R}_{ee}}$ based on the polymer end-to-end vector $\vec{R}_{ee}$.  For the simulations of $Z_k=24.8$ \cite{nek-19a,sek-19} we approximate $\tens{S}^B$ using  the orientation tensor  computed by \citet{sek-19} from the unit vectors pointing between adjacent kinks from the Z1 code.  Both cases lead to an approximate value of $\beta^B=2.0$, which is larger than $\beta^{\textrm{A}}=\pea\pm\pead$ found using Method A.  We know that both approximations will overestimate the tube segment orientation.  The degree of orientation of the end-to-end vector exceeds the degree of orientation of \change{shorter sub-}segments \delete{that connect that end-to-end vector} (except for the rare case of a fully-stretched  polymer), which would lead to a larger fitted value of $\beta$ in order to match the simulations. Similarly, the orientation computed in the Z1 code exceeds that computed from the full chain because the Z1 code averages out fluctuations in entanglement positions, which again leads to a larger extracted value for $\beta$.

\begin{figure}
	\centering
	\includegraphics[width=0.95\textwidth]{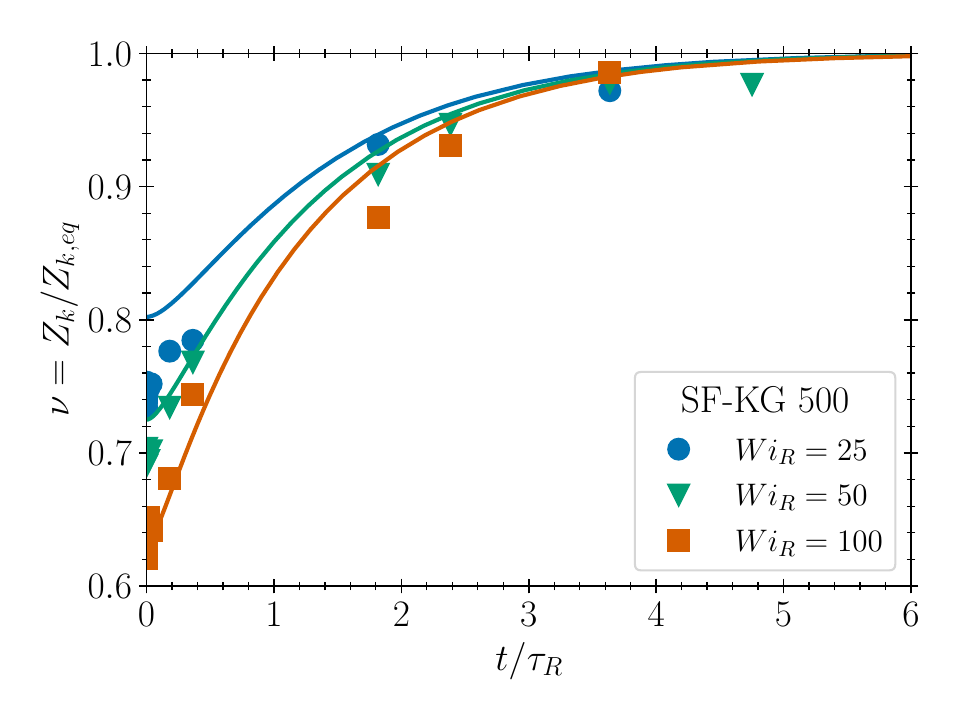}
	\caption{Re-entanglement following cessation of steady-shear. Filled circles are SF-KG 500 simulations, solid lines represent numerical solutions of Eq.~\eqref{eq:governing-equations}  
 using the value $\beta=\sfl$ from fitting to the steady state shear data using Method A (Figure~\ref{fig:Marco-steady-state}).
 }\label{fig:cessation}
\end{figure}

We now examine the kinetic predictions of our model by comparing numerical solutions of Eqs.~\eqref{eq:governing-equations} to the time-dependent re-entanglement in SF-KG chains following cessation of steady shear flow.  We use the value $\beta=\sfl$ obtained from Method A. Results of  molecular dynamics simulations are compared with model predictions in Figure~\ref{fig:cessation} for three values of the Rouse Weissenberg number (${\Wi}_R=25,50,100$).  We observe good agreement between theory and simulation, with both displaying rapid re-entanglement by $3\tau_R$, which confirms earlier suggestions that the melt re-entangles on the Rouse time.\cite{ohr-19,bnek-22,dolata2023} 
 \textit{i.e.}, before the stress fully relaxes.  Physically, the stretch and number of entanglements  return to their equilibrium value on the Rouse time due to chain retraction in the still oriented tube. The subsequent relaxation of orientation requires renewal of the tube, which can only happen on the reptation time.  This physical picture is consistent with molecular dynamics simulations,\cite{ohr-19} and contrasts with the interpretation of experiments that claim to  determine entanglement dynamics from the rheology of interrupted shear \delete{experiments}.\cite{rr-13} Further results for re-entanglement and steady state rheology are given in the Supplementary Information.

The fast re-entanglement within a time $\sim\tau_R$ suggests that experiments that rely on stress measurements cannot unambiguously determine the state of entanglement. A common method that is claimed to measure reentanglement is to perform repeated shear startup experiments, where the second startup occurs a time $\tau_{\textrm{wait}}$ before the first experiment has fully relaxed after cessation of flow.\cite{rr-13,wang2013new,galvani2022probing} For a short $\tau_{\textrm{wait}}<\tau_d$ the \delete{second} stress overshoot \change{during the second startup} is very weak, which has been interpreted as a reduction of entanglements. It typically takes a relaxation time of order the reptation time, or even longer, for the second overshoot to reproduce that of a fully entangled melt \cite{rr-13,wang2013new,galvani2022probing}. A large contribution to the decrease in the second stress overshoot in those experiments is probably  due to the slow relaxation of chain alignment via reptation, rather than slow recovery of entanglements.\cite{im-14c}. Simulations by  \citeauthor{galvani2022probing}\cite{galvani2022probing} showed that entanglements fully recover on the Rouse time. Moreover, these dynamics can also be crudely captured by the Rolie-Poly model, which assumes no change in entanglement number\cite{GHO-MM2013}. As a consequence, the second stress overshoot should not be interpreted in terms of entanglement recovery. Similar behavior was observed in simulations of the  welding of flow aligned layers, where the strength of re-entangled ``healed'' layers \change{was} degraded by residual alignment at the weld location, rather than by a loss of entanglements\cite{cr-20}.

In summary, we have extracted the CCR parameter $\beta$ by fitting a constitutive equation and a structural model to steady-state simulations of KG and UA-PE melts. We have shown that $\beta$ is independent of molecular weight but depends on polymer `chemistry', with $\beta$ scaling approximately as $(b_K/p)^{1.9}$,
which implies that stiffer melts with larger tube diameters will disentangle more  readily than more extensible melts.  Furthermore, re-entanglement of the melt on the Rouse time was confirmed. Finally, we note that the magnitude of $\beta$ is believed to influence shear banding behaviour in entangled polymers \cite{adams2011transient}. In the absence of CCR ($\beta=0$) the shear stress has a maximum as a function of shear rate, which implies an instability to shear banding. Our results suggest that polymers with larger $\beta$, corresponding to larger $b_K/p$, could be less susceptible to shear banding \cite{adams2011transient}.

\begin{acknowledgement}
The authors thank Mark O. Robbins, Jon E. Seppala, Gretar Tryggvason, and Thao (Vicky) Nguyen for useful discussions during the development of the model.  This work was funded in part by NSF DMREF award 1628974. PDO is grateful to Georgetown University and the Ives foundation for support. We thank Martin Kr\"oger for a copy of the Z1 code.

Certain commercial or open-source software are identified in this paper in order to specify the methodology adequately. Such identification is not intended to imply recommendation or endorsement of any product or service by NIST, nor is it intended to imply that the software identified are necessarily the best available for the purpose.

\end{acknowledgement}
\begin{suppinfo}
The following information is available free of charge:

\change{Dolata\_SI\_Final.pdf} 

\noindent It describes parameter determination, simulation details, and more comparison with the Ianniruberto-Marrucci model.
\end{suppinfo}

\begin{tocentry}
\includegraphics[width=\textwidth,angle=-90]{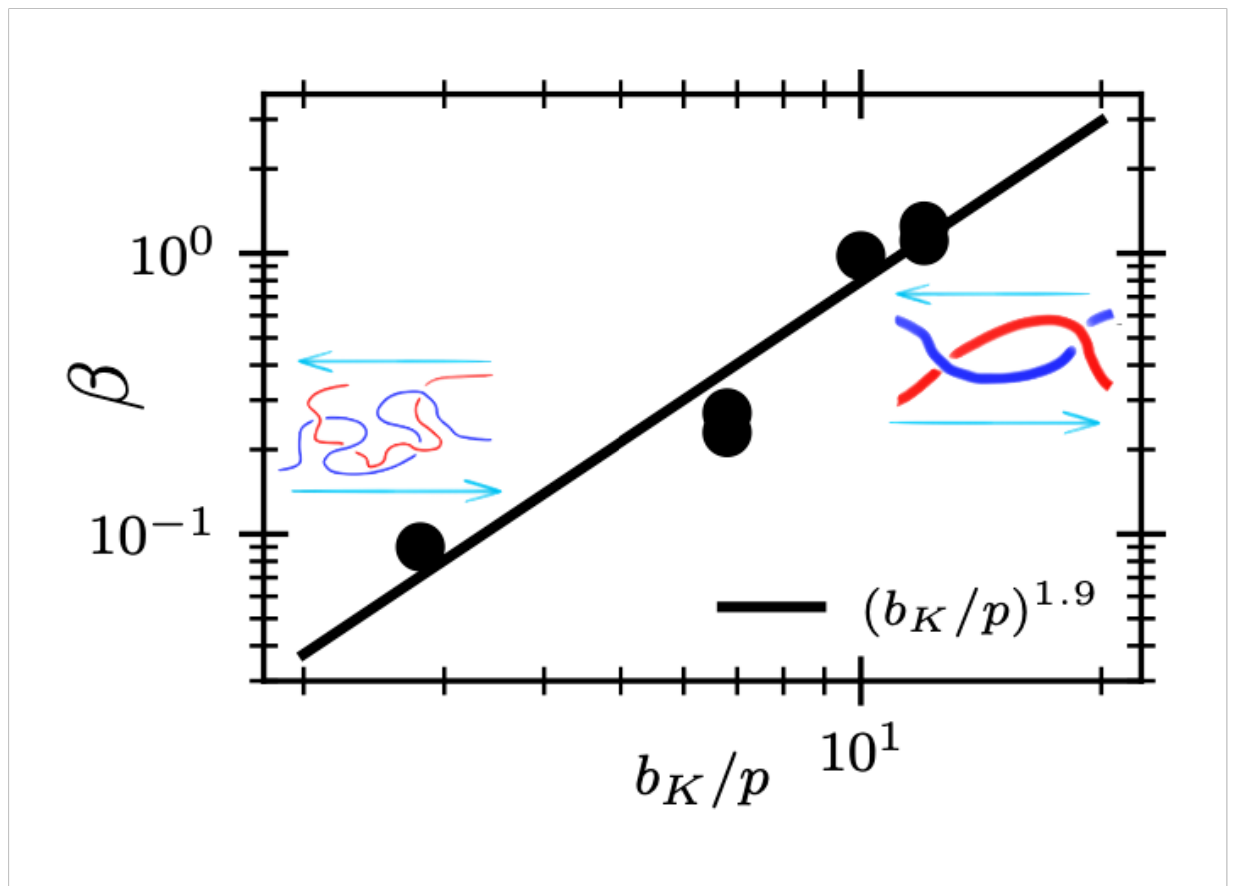}
CCR parameter as a function of polymer chain characteristics.
\end{tocentry}
\bibliography{GENERIC-ref.bib}

\providecommand{\latin}[1]{#1}
\makeatletter
\providecommand{\doi}
  {\begingroup\let\do\@makeother\dospecials
  \catcode`\{=1 \catcode`\}=2 \doi@aux}
\providecommand{\doi@aux}[1]{\endgroup\texttt{#1}}
\makeatother
\providecommand*\mcitethebibliography{\thebibliography}
\csname @ifundefined\endcsname{endmcitethebibliography}
  {\let\endmcitethebibliography\endthebibliography}{}
\begin{mcitethebibliography}{50}
\providecommand*\natexlab[1]{#1}
\providecommand*\mciteSetBstSublistMode[1]{}
\providecommand*\mciteSetBstMaxWidthForm[2]{}
\providecommand*\mciteBstWouldAddEndPuncttrue
  {\def\EndOfBibitem{\unskip.}}
\providecommand*\mciteBstWouldAddEndPunctfalse
  {\let\EndOfBibitem\relax}
\providecommand*\mciteSetBstMidEndSepPunct[3]{}
\providecommand*\mciteSetBstSublistLabelBeginEnd[3]{}
\providecommand*\EndOfBibitem{}
\mciteSetBstSublistMode{f}
\mciteSetBstMaxWidthForm{subitem}{(\alph{mcitesubitemcount})}
\mciteSetBstSublistLabelBeginEnd
  {\mcitemaxwidthsubitemform\space}
  {\relax}
  {\relax}

\bibitem[Doi and Edwards(1988)Doi, and Edwards]{DoiEdwards_Book}
Doi,~M.; Edwards,~S.~F. \emph{The Theory of Polymer Dynamics}; Oxford
  University Press, 1988\relax
\mciteBstWouldAddEndPuncttrue
\mciteSetBstMidEndSepPunct{\mcitedefaultmidpunct}
{\mcitedefaultendpunct}{\mcitedefaultseppunct}\relax
\EndOfBibitem
\bibitem[Masubuchi \latin{et~al.}(2003)Masubuchi, Ianniruberto, Greco, and
  Marrucci]{maigm-03}
Masubuchi,~Y.; Ianniruberto,~G.; Greco,~F.; Marrucci,~G. \emph{The Journal of
  Chemical Physics} \textbf{2003}, \emph{119}, 6925--6930\relax
\mciteBstWouldAddEndPuncttrue
\mciteSetBstMidEndSepPunct{\mcitedefaultmidpunct}
{\mcitedefaultendpunct}{\mcitedefaultseppunct}\relax
\EndOfBibitem
\bibitem[Foteinopoulou \latin{et~al.}(2006)Foteinopoulou, Karayiannis,
  Mavrantzas, and Kr{\"o}ger]{fkmk-06}
Foteinopoulou,~K.; Karayiannis,~N.~C.; Mavrantzas,~V.~G.; Kr{\"o}ger,~M.
  \emph{Macromolecules} \textbf{2006}, \emph{39}, 4207--4216\relax
\mciteBstWouldAddEndPuncttrue
\mciteSetBstMidEndSepPunct{\mcitedefaultmidpunct}
{\mcitedefaultendpunct}{\mcitedefaultseppunct}\relax
\EndOfBibitem
\bibitem[Baig \latin{et~al.}(2010)Baig, Mavrantzas, and Kr{\"o}ger]{bmk-10}
Baig,~C.; Mavrantzas,~V.~G.; Kr{\"o}ger,~M. \emph{Macromolecules}
  \textbf{2010}, \emph{43}, 6886--6902\relax
\mciteBstWouldAddEndPuncttrue
\mciteSetBstMidEndSepPunct{\mcitedefaultmidpunct}
{\mcitedefaultendpunct}{\mcitedefaultseppunct}\relax
\EndOfBibitem
\bibitem[Nafar~Sefiddashti \latin{et~al.}(2015)Nafar~Sefiddashti, Edwards, and
  Khomami]{sek-15}
Nafar~Sefiddashti,~M.~H.; Edwards,~B.~J.; Khomami,~B. \emph{Journal of
  Rheology} \textbf{2015}, \emph{59}, 119--153\relax
\mciteBstWouldAddEndPuncttrue
\mciteSetBstMidEndSepPunct{\mcitedefaultmidpunct}
{\mcitedefaultendpunct}{\mcitedefaultseppunct}\relax
\EndOfBibitem
\bibitem[Nafar~Sefiddashti \latin{et~al.}(2016)Nafar~Sefiddashti, Edwards, and
  Khomami]{sek-16}
Nafar~Sefiddashti,~M.~H.; Edwards,~B.~J.; Khomami,~B. \emph{Journal of
  Rheology} \textbf{2016}, \emph{60}, 1227--1244\relax
\mciteBstWouldAddEndPuncttrue
\mciteSetBstMidEndSepPunct{\mcitedefaultmidpunct}
{\mcitedefaultendpunct}{\mcitedefaultseppunct}\relax
\EndOfBibitem
\bibitem[Nafar~Sefiddashti \latin{et~al.}(2019)Nafar~Sefiddashti, Edwards, and
  Khomami]{nek-19a}
Nafar~Sefiddashti,~M.~H.; Edwards,~B.~J.; Khomami,~B. \emph{Polymers}
  \textbf{2019}, \emph{11}, 476\relax
\mciteBstWouldAddEndPuncttrue
\mciteSetBstMidEndSepPunct{\mcitedefaultmidpunct}
{\mcitedefaultendpunct}{\mcitedefaultseppunct}\relax
\EndOfBibitem
\bibitem[Nafar~Sefiddashti \latin{et~al.}(2019)Nafar~Sefiddashti, Edwards, and
  Khomami]{sek-19}
Nafar~Sefiddashti,~M.~H.; Edwards,~B.~J.; Khomami,~B. \emph{Macromolecules}
  \textbf{2019}, \emph{52}, 8124--8143\relax
\mciteBstWouldAddEndPuncttrue
\mciteSetBstMidEndSepPunct{\mcitedefaultmidpunct}
{\mcitedefaultendpunct}{\mcitedefaultseppunct}\relax
\EndOfBibitem
\bibitem[Galvani~Cunha \latin{et~al.}(2022)Galvani~Cunha, Olmsted, and
  Robbins]{galvani2022probing}
Galvani~Cunha,~M.~A.; Olmsted,~P.~D.; Robbins,~M.~O. \emph{Journal of Rheology}
  \textbf{2022}, \emph{66}, 619--637\relax
\mciteBstWouldAddEndPuncttrue
\mciteSetBstMidEndSepPunct{\mcitedefaultmidpunct}
{\mcitedefaultendpunct}{\mcitedefaultseppunct}\relax
\EndOfBibitem
\bibitem[Graham \latin{et~al.}(2003)Graham, Likhtman, McLeish, and
  Milner]{glmm-03}
Graham,~R.~S.; Likhtman,~A.~E.; McLeish,~T. C.~B.; Milner,~S.~T. \emph{Journal
  of Rheology} \textbf{2003}, \emph{47}, 1171--1200\relax
\mciteBstWouldAddEndPuncttrue
\mciteSetBstMidEndSepPunct{\mcitedefaultmidpunct}
{\mcitedefaultendpunct}{\mcitedefaultseppunct}\relax
\EndOfBibitem
\bibitem[Marrucci(1996)]{marrucci-96}
Marrucci,~G. \emph{Journal of Non-Newtonian Fluid Mechanics} \textbf{1996},
  \emph{62}, 279--289\relax
\mciteBstWouldAddEndPuncttrue
\mciteSetBstMidEndSepPunct{\mcitedefaultmidpunct}
{\mcitedefaultendpunct}{\mcitedefaultseppunct}\relax
\EndOfBibitem
\bibitem[Ianniruberto and Marrucci(2014)Ianniruberto, and Marrucci]{im-14}
Ianniruberto,~G.; Marrucci,~G. \emph{Journal of Rheology} \textbf{2014},
  \emph{58}, 89--102\relax
\mciteBstWouldAddEndPuncttrue
\mciteSetBstMidEndSepPunct{\mcitedefaultmidpunct}
{\mcitedefaultendpunct}{\mcitedefaultseppunct}\relax
\EndOfBibitem
\bibitem[Ianniruberto and Marrucci(2014)Ianniruberto, and Marrucci]{im-14b}
Ianniruberto,~G.; Marrucci,~G. \emph{Journal of Rheology} \textbf{2014},
  \emph{58}, 1083--1083\relax
\mciteBstWouldAddEndPuncttrue
\mciteSetBstMidEndSepPunct{\mcitedefaultmidpunct}
{\mcitedefaultendpunct}{\mcitedefaultseppunct}\relax
\EndOfBibitem
\bibitem[Hawke \latin{et~al.}(2015)Hawke, Huang, Hassager, and Read]{hhhr-15}
Hawke,~L. G.~D.; Huang,~Q.; Hassager,~O.; Read,~D.~J. \emph{Journal of
  Rheology} \textbf{2015}, \emph{59}, 995--1017\relax
\mciteBstWouldAddEndPuncttrue
\mciteSetBstMidEndSepPunct{\mcitedefaultmidpunct}
{\mcitedefaultendpunct}{\mcitedefaultseppunct}\relax
\EndOfBibitem
\bibitem[Dolata and Olmsted(2023)Dolata, and Olmsted]{dolata2023}
Dolata,~B.~E.; Olmsted,~P.~D. \emph{Journal of Rheology} \textbf{2023},
  \emph{67}, 269--292\relax
\mciteBstWouldAddEndPuncttrue
\mciteSetBstMidEndSepPunct{\mcitedefaultmidpunct}
{\mcitedefaultendpunct}{\mcitedefaultseppunct}\relax
\EndOfBibitem
\bibitem[Tapadia and Wang(2004)Tapadia, and Wang]{Tapadia2004Nonlinear-flow-}
Tapadia,~P.; Wang,~S.-Q. \emph{Macromolecules} \textbf{2004}, \emph{37},
  9083--9095\relax
\mciteBstWouldAddEndPuncttrue
\mciteSetBstMidEndSepPunct{\mcitedefaultmidpunct}
{\mcitedefaultendpunct}{\mcitedefaultseppunct}\relax
\EndOfBibitem
\bibitem[Sanchez-Reyes and Archer(2002)Sanchez-Reyes, and Archer]{sa-02}
Sanchez-Reyes,~J.; Archer,~L. \emph{Macromolecules} \textbf{2002}, \emph{35},
  5194--5202\relax
\mciteBstWouldAddEndPuncttrue
\mciteSetBstMidEndSepPunct{\mcitedefaultmidpunct}
{\mcitedefaultendpunct}{\mcitedefaultseppunct}\relax
\EndOfBibitem
\bibitem[Roy and Roland(2013)Roy, and Roland]{rr-13}
Roy,~D.; Roland,~C.~M. \emph{Macromolecules} \textbf{2013}, \emph{46},
  9403--9408\relax
\mciteBstWouldAddEndPuncttrue
\mciteSetBstMidEndSepPunct{\mcitedefaultmidpunct}
{\mcitedefaultendpunct}{\mcitedefaultseppunct}\relax
\EndOfBibitem
\bibitem[Xu \latin{et~al.}(2022)Xu, Sun, Lu, Patil, Mays, Schweizer, and
  Cheng]{xu-22}
Xu,~Z.; Sun,~R.; Lu,~W.; Patil,~S.; Mays,~J.; Schweizer,~K.; Cheng,~S.
  \emph{Macromolecules} \textbf{2022}, \emph{55}, 10737--10750\relax
\mciteBstWouldAddEndPuncttrue
\mciteSetBstMidEndSepPunct{\mcitedefaultmidpunct}
{\mcitedefaultendpunct}{\mcitedefaultseppunct}\relax
\EndOfBibitem
\bibitem[Wool \latin{et~al.}(1989)Wool, Yuan, and McGarel]{wym-89}
Wool,~R.~P.; Yuan,~B.-L.; McGarel,~O. \emph{Polymer Engineering \& Science}
  \textbf{1989}, \emph{29}, 1340--1367\relax
\mciteBstWouldAddEndPuncttrue
\mciteSetBstMidEndSepPunct{\mcitedefaultmidpunct}
{\mcitedefaultendpunct}{\mcitedefaultseppunct}\relax
\EndOfBibitem
\bibitem[O'Connor and McLeish(1993)O'Connor, and McLeish]{om-93}
O'Connor,~K.; McLeish,~T. \emph{Macromolecules} \textbf{1993}, \emph{26},
  7322--7325\relax
\mciteBstWouldAddEndPuncttrue
\mciteSetBstMidEndSepPunct{\mcitedefaultmidpunct}
{\mcitedefaultendpunct}{\mcitedefaultseppunct}\relax
\EndOfBibitem
\bibitem[Ge \latin{et~al.}(2013)Ge, Pierce, Perahia, Grest, and
  Robbins]{gppgr-13}
Ge,~T.; Pierce,~F.; Perahia,~D.; Grest,~G.~S.; Robbins,~M.~O. \emph{Phys. Rev.
  Lett.} \textbf{2013}, \emph{110}, 098301\relax
\mciteBstWouldAddEndPuncttrue
\mciteSetBstMidEndSepPunct{\mcitedefaultmidpunct}
{\mcitedefaultendpunct}{\mcitedefaultseppunct}\relax
\EndOfBibitem
\bibitem[Cunha and Robbins(2020)Cunha, and Robbins]{cr-20}
Cunha,~M. A.~G.; Robbins,~M.~O. \emph{Macromolecules} \textbf{2020}, \emph{53},
  8417--8427\relax
\mciteBstWouldAddEndPuncttrue
\mciteSetBstMidEndSepPunct{\mcitedefaultmidpunct}
{\mcitedefaultendpunct}{\mcitedefaultseppunct}\relax
\EndOfBibitem
\bibitem[Kremer and Grest(1990)Kremer, and Grest]{kg-90}
Kremer,~K.; Grest,~G.~S. \emph{The Journal of Chemical Physics} \textbf{1990},
  \emph{92}, 5057--5086\relax
\mciteBstWouldAddEndPuncttrue
\mciteSetBstMidEndSepPunct{\mcitedefaultmidpunct}
{\mcitedefaultendpunct}{\mcitedefaultseppunct}\relax
\EndOfBibitem
\bibitem[Ianniruberto(2015)]{ianniruberto-15}
Ianniruberto,~G. \emph{Journal of Rheology} \textbf{2015}, \emph{59},
  211--235\relax
\mciteBstWouldAddEndPuncttrue
\mciteSetBstMidEndSepPunct{\mcitedefaultmidpunct}
{\mcitedefaultendpunct}{\mcitedefaultseppunct}\relax
\EndOfBibitem
\bibitem[Kr{\"o}ger(2005)]{kroger-05}
Kr{\"o}ger,~M. \emph{Computer Physics Communications} \textbf{2005},
  \emph{168}, 209--232\relax
\mciteBstWouldAddEndPuncttrue
\mciteSetBstMidEndSepPunct{\mcitedefaultmidpunct}
{\mcitedefaultendpunct}{\mcitedefaultseppunct}\relax
\EndOfBibitem
\bibitem[Tzoumanekas and Theodorou(2006)Tzoumanekas, and Theodorou]{tt-06}
Tzoumanekas,~C.; Theodorou,~D.~N. \emph{Macromolecules} \textbf{2006},
  \emph{39}, 4592--4604\relax
\mciteBstWouldAddEndPuncttrue
\mciteSetBstMidEndSepPunct{\mcitedefaultmidpunct}
{\mcitedefaultendpunct}{\mcitedefaultseppunct}\relax
\EndOfBibitem
\bibitem[Likhtman and Graham(2003)Likhtman, and Graham]{lg-03}
Likhtman,~A.~E.; Graham,~R.~S. \emph{Journal of Non-Newtonian Fluid Mechanics}
  \textbf{2003}, \emph{114}, 1--12\relax
\mciteBstWouldAddEndPuncttrue
\mciteSetBstMidEndSepPunct{\mcitedefaultmidpunct}
{\mcitedefaultendpunct}{\mcitedefaultseppunct}\relax
\EndOfBibitem
\bibitem[Likhtman and McLeish(2002)Likhtman, and McLeish]{lm-02}
Likhtman,~A.~E.; McLeish,~T. C.~B. \emph{Macromolecules} \textbf{2002},
  \emph{35}, 6332--6343\relax
\mciteBstWouldAddEndPuncttrue
\mciteSetBstMidEndSepPunct{\mcitedefaultmidpunct}
{\mcitedefaultendpunct}{\mcitedefaultseppunct}\relax
\EndOfBibitem
\bibitem[Cohen(1991)]{cohen-91}
Cohen,~A. \emph{Rheological Acta} \textbf{1991}, \emph{30}, 270--273\relax
\mciteBstWouldAddEndPuncttrue
\mciteSetBstMidEndSepPunct{\mcitedefaultmidpunct}
{\mcitedefaultendpunct}{\mcitedefaultseppunct}\relax
\EndOfBibitem
\bibitem[Stephanou \latin{et~al.}(2009)Stephanou, Baig, and Mavrantzas]{sbm-09}
Stephanou,~P.~S.; Baig,~C.; Mavrantzas,~V.~G. \emph{Journal of Rheology}
  \textbf{2009}, \emph{53}, 309--337\relax
\mciteBstWouldAddEndPuncttrue
\mciteSetBstMidEndSepPunct{\mcitedefaultmidpunct}
{\mcitedefaultendpunct}{\mcitedefaultseppunct}\relax
\EndOfBibitem
\bibitem[Mead \latin{et~al.}(2015)Mead, Banerjee, and Park]{mbp-15}
Mead,~D.~W.; Banerjee,~N.; Park,~J. \emph{Journal of Rheology} \textbf{2015},
  \emph{59}, 335--363\relax
\mciteBstWouldAddEndPuncttrue
\mciteSetBstMidEndSepPunct{\mcitedefaultmidpunct}
{\mcitedefaultendpunct}{\mcitedefaultseppunct}\relax
\EndOfBibitem
\bibitem[Mead \latin{et~al.}(2018)Mead, Monjezi, and Park]{mmp-18}
Mead,~D.~W.; Monjezi,~S.; Park,~J. \emph{Journal of Rheology} \textbf{2018},
  \emph{62}, 121--134\relax
\mciteBstWouldAddEndPuncttrue
\mciteSetBstMidEndSepPunct{\mcitedefaultmidpunct}
{\mcitedefaultendpunct}{\mcitedefaultseppunct}\relax
\EndOfBibitem
\bibitem[O'Connor \latin{et~al.}(2019)O'Connor, Hopkins, and Robbins]{ohr-19}
O'Connor,~T.~C.; Hopkins,~A.; Robbins,~M.~O. \emph{Macromolecules}
  \textbf{2019}, \emph{52}, 8540--8550\relax
\mciteBstWouldAddEndPuncttrue
\mciteSetBstMidEndSepPunct{\mcitedefaultmidpunct}
{\mcitedefaultendpunct}{\mcitedefaultseppunct}\relax
\EndOfBibitem
\bibitem[Qin(2024)]{qin-24}
Qin,~J. \emph{Macromolecules} \textbf{2024}, \relax
\mciteBstWouldAddEndPunctfalse
\mciteSetBstMidEndSepPunct{\mcitedefaultmidpunct}
{}{\mcitedefaultseppunct}\relax
\EndOfBibitem
\bibitem[Hoy and Kr{\"o}ger(2020)Hoy, and Kr{\"o}ger]{hk-20}
Hoy,~R.~S.; Kr{\"o}ger,~M. \emph{Physical Review Letters} \textbf{2020},
  \emph{124}, 147801\relax
\mciteBstWouldAddEndPuncttrue
\mciteSetBstMidEndSepPunct{\mcitedefaultmidpunct}
{\mcitedefaultendpunct}{\mcitedefaultseppunct}\relax
\EndOfBibitem
\bibitem[Milner(2020)]{milner-20}
Milner,~S.~T. \emph{Macromolecules} \textbf{2020}, \emph{53}, 1314--1325\relax
\mciteBstWouldAddEndPuncttrue
\mciteSetBstMidEndSepPunct{\mcitedefaultmidpunct}
{\mcitedefaultendpunct}{\mcitedefaultseppunct}\relax
\EndOfBibitem
\bibitem[Dietz \latin{et~al.}(2022)Dietz, Kr{\"o}ger, and Hoy]{dkh-22}
Dietz,~J.~D.; Kr{\"o}ger,~M.; Hoy,~R.~S. \emph{Macromolecules} \textbf{2022},
  \emph{55}, 3613--3626\relax
\mciteBstWouldAddEndPuncttrue
\mciteSetBstMidEndSepPunct{\mcitedefaultmidpunct}
{\mcitedefaultendpunct}{\mcitedefaultseppunct}\relax
\EndOfBibitem
\bibitem[Auhl \latin{et~al.}(2003)Auhl, Everaers, Grest, Kremer, and
  Plimpton]{auhl2003equilibration}
Auhl,~R.; Everaers,~R.; Grest,~G.~S.; Kremer,~K.; Plimpton,~S.~J. \emph{The
  Journal of Chemical Physics} \textbf{2003}, \emph{119}, 12718--12728\relax
\mciteBstWouldAddEndPuncttrue
\mciteSetBstMidEndSepPunct{\mcitedefaultmidpunct}
{\mcitedefaultendpunct}{\mcitedefaultseppunct}\relax
\EndOfBibitem
\bibitem[Everaers \latin{et~al.}(2020)Everaers, Karimi-Varzaneh, Fleck, Hojdis,
  and Svaneborg]{ekfhs-20}
Everaers,~R.; Karimi-Varzaneh,~H.~A.; Fleck,~F.; Hojdis,~N.; Svaneborg,~C.
  \emph{Macromolecules} \textbf{2020}, \emph{53}, 1901--1916\relax
\mciteBstWouldAddEndPuncttrue
\mciteSetBstMidEndSepPunct{\mcitedefaultmidpunct}
{\mcitedefaultendpunct}{\mcitedefaultseppunct}\relax
\EndOfBibitem
\bibitem[Foteinopoulou \latin{et~al.}(2009)Foteinopoulou, Karayiannis, Laso,
  and Kro{\"o}ger]{fklk-09}
Foteinopoulou,~K.; Karayiannis,~N.~C.; Laso,~M.; Kro{\"o}ger,~M. \emph{J. of
  Phys. Chem. B} \textbf{2009}, \emph{113}, 442--455\relax
\mciteBstWouldAddEndPuncttrue
\mciteSetBstMidEndSepPunct{\mcitedefaultmidpunct}
{\mcitedefaultendpunct}{\mcitedefaultseppunct}\relax
\EndOfBibitem
\bibitem[Sefiddashti \latin{et~al.}(2018)Sefiddashti, Edwards, and
  Khomami]{sek-18extension}
Sefiddashti,~M. H.~N.; Edwards,~B.~J.; Khomami,~B. \emph{Physical Review
  Letters} \textbf{2018}, \emph{121}, 247802\relax
\mciteBstWouldAddEndPuncttrue
\mciteSetBstMidEndSepPunct{\mcitedefaultmidpunct}
{\mcitedefaultendpunct}{\mcitedefaultseppunct}\relax
\EndOfBibitem
\bibitem[Mendes~Jr \latin{et~al.}(1991)Mendes~Jr, Lindner, Buzier, Boue, and
  Bastide]{mendes1991experimental}
Mendes~Jr,~E.; Lindner,~P.; Buzier,~M.; Boue,~F.; Bastide,~J. \emph{Physical
  Review Letters} \textbf{1991}, \emph{66}, 1595\relax
\mciteBstWouldAddEndPuncttrue
\mciteSetBstMidEndSepPunct{\mcitedefaultmidpunct}
{\mcitedefaultendpunct}{\mcitedefaultseppunct}\relax
\EndOfBibitem
\bibitem[Hsu and Kremer(2018)Hsu, and Kremer]{hk-18}
Hsu,~H.-P.; Kremer,~K. \emph{ACS Macro Letters} \textbf{2018}, \emph{7},
  107--111\relax
\mciteBstWouldAddEndPuncttrue
\mciteSetBstMidEndSepPunct{\mcitedefaultmidpunct}
{\mcitedefaultendpunct}{\mcitedefaultseppunct}\relax
\EndOfBibitem
\bibitem[Boudaghi \latin{et~al.}(2022)Boudaghi, Nafar~Seddashti, Edwards, and
  Khomami]{bnek-22}
Boudaghi,~M.; Nafar~Seddashti,~M.~H.; Edwards,~B.~J.; Khomami,~B. \emph{Journal
  of Rheology} \textbf{2022}, \emph{66}, 551--569\relax
\mciteBstWouldAddEndPuncttrue
\mciteSetBstMidEndSepPunct{\mcitedefaultmidpunct}
{\mcitedefaultendpunct}{\mcitedefaultseppunct}\relax
\EndOfBibitem
\bibitem[Wang \latin{et~al.}(2013)Wang, Wang, Cheng, Li, Zhu, and
  Sun]{wang2013new}
Wang,~S.-Q.; Wang,~Y.; Cheng,~S.; Li,~X.; Zhu,~X.; Sun,~H.
  \emph{Macromolecules} \textbf{2013}, \emph{46}, 3147--3159\relax
\mciteBstWouldAddEndPuncttrue
\mciteSetBstMidEndSepPunct{\mcitedefaultmidpunct}
{\mcitedefaultendpunct}{\mcitedefaultseppunct}\relax
\EndOfBibitem
\bibitem[Ianniruberto and Marrucci(2014)Ianniruberto, and Marrucci]{im-14c}
Ianniruberto,~G.; Marrucci,~G. \emph{ACS Macro Letters} \textbf{2014},
  \emph{3}, 552--555\relax
\mciteBstWouldAddEndPuncttrue
\mciteSetBstMidEndSepPunct{\mcitedefaultmidpunct}
{\mcitedefaultendpunct}{\mcitedefaultseppunct}\relax
\EndOfBibitem
\bibitem[Graham \latin{et~al.}(2013)Graham, Henry, and Olmsted]{GHO-MM2013}
Graham,~R.~S.; Henry,~E.~P.; Olmsted,~P.~D. \emph{Macromolecules}
  \textbf{2013}, \emph{46}, 9849--9854\relax
\mciteBstWouldAddEndPuncttrue
\mciteSetBstMidEndSepPunct{\mcitedefaultmidpunct}
{\mcitedefaultendpunct}{\mcitedefaultseppunct}\relax
\EndOfBibitem
\bibitem[Adams \latin{et~al.}(2011)Adams, Fielding, and
  Olmsted]{adams2011transient}
Adams,~J.; Fielding,~S.~M.; Olmsted,~P.~D. \emph{Journal of Rheology}
  \textbf{2011}, \emph{55}, 1007--1032\relax
\mciteBstWouldAddEndPuncttrue
\mciteSetBstMidEndSepPunct{\mcitedefaultmidpunct}
{\mcitedefaultendpunct}{\mcitedefaultseppunct}\relax
\EndOfBibitem
\end{mcitethebibliography}


\providecommand{\latin}[1]{#1}
\makeatletter
\providecommand{\doi}
  {\begingroup\let\do\@makeother\dospecials
  \catcode`\{=1 \catcode`\}=2 \doi@aux}
\providecommand{\doi@aux}[1]{\endgroup\texttt{#1}}
\makeatother
\providecommand*\mcitethebibliography{\thebibliography}
\csname @ifundefined\endcsname{endmcitethebibliography}
  {\let\endmcitethebibliography\endthebibliography}{}
\begin{mcitethebibliography}{29}
\providecommand*\natexlab[1]{#1}
\providecommand*\mciteSetBstSublistMode[1]{}
\providecommand*\mciteSetBstMaxWidthForm[2]{}
\providecommand*\mciteBstWouldAddEndPuncttrue
  {\def\EndOfBibitem{\unskip.}}
\providecommand*\mciteBstWouldAddEndPunctfalse
  {\let\EndOfBibitem\relax}
\providecommand*\mciteSetBstMidEndSepPunct[3]{}
\providecommand*\mciteSetBstSublistLabelBeginEnd[3]{}
\providecommand*\EndOfBibitem{}
\mciteSetBstSublistMode{f}
\mciteSetBstMaxWidthForm{subitem}{(\alph{mcitesubitemcount})}
\mciteSetBstSublistLabelBeginEnd
  {\mcitemaxwidthsubitemform\space}
  {\relax}
  {\relax}

\bibitem[Shanbhag and Kr{\"o}ger(2007)Shanbhag, and Kr{\"o}ger]{sk-07}
Shanbhag,~S.; Kr{\"o}ger,~M. \emph{Macromolecules} \textbf{2007}, \emph{40},
  2897--2903\relax
\mciteBstWouldAddEndPuncttrue
\mciteSetBstMidEndSepPunct{\mcitedefaultmidpunct}
{\mcitedefaultendpunct}{\mcitedefaultseppunct}\relax
\EndOfBibitem
\bibitem[Dolata and Olmsted(2023)Dolata, and Olmsted]{dolata2023}
Dolata,~B.~E.; Olmsted,~P.~D. \emph{Journal of Rheology} \textbf{2023},
  \emph{67}, 269--292\relax
\mciteBstWouldAddEndPuncttrue
\mciteSetBstMidEndSepPunct{\mcitedefaultmidpunct}
{\mcitedefaultendpunct}{\mcitedefaultseppunct}\relax
\EndOfBibitem
\bibitem[Cohen(1991)]{cohen-91}
Cohen,~A. \emph{Rheological Acta} \textbf{1991}, \emph{30}, 270--273\relax
\mciteBstWouldAddEndPuncttrue
\mciteSetBstMidEndSepPunct{\mcitedefaultmidpunct}
{\mcitedefaultendpunct}{\mcitedefaultseppunct}\relax
\EndOfBibitem
\bibitem[Stephanou \latin{et~al.}(2009)Stephanou, Baig, and Mavrantzas]{sbm-09}
Stephanou,~P.~S.; Baig,~C.; Mavrantzas,~V.~G. \emph{Journal of Rheology}
  \textbf{2009}, \emph{53}, 309--337\relax
\mciteBstWouldAddEndPuncttrue
\mciteSetBstMidEndSepPunct{\mcitedefaultmidpunct}
{\mcitedefaultendpunct}{\mcitedefaultseppunct}\relax
\EndOfBibitem
\bibitem[Everaers \latin{et~al.}(2020)Everaers, Karimi-Varzaneh, Fleck, Hojdis,
  and Svaneborg]{ekfhs-20}
Everaers,~R.; Karimi-Varzaneh,~H.~A.; Fleck,~F.; Hojdis,~N.; Svaneborg,~C.
  \emph{Macromolecules} \textbf{2020}, \emph{53}, 1901--1916\relax
\mciteBstWouldAddEndPuncttrue
\mciteSetBstMidEndSepPunct{\mcitedefaultmidpunct}
{\mcitedefaultendpunct}{\mcitedefaultseppunct}\relax
\EndOfBibitem
\bibitem[Nafar~Sefiddashti \latin{et~al.}(2015)Nafar~Sefiddashti, Edwards, and
  Khomami]{sek-15}
Nafar~Sefiddashti,~M.~H.; Edwards,~B.~J.; Khomami,~B. \emph{Journal of
  Rheology} \textbf{2015}, \emph{59}, 119--153\relax
\mciteBstWouldAddEndPuncttrue
\mciteSetBstMidEndSepPunct{\mcitedefaultmidpunct}
{\mcitedefaultendpunct}{\mcitedefaultseppunct}\relax
\EndOfBibitem
\bibitem[Nafar~Sefiddashti \latin{et~al.}(2016)Nafar~Sefiddashti, Edwards, and
  Khomami]{sek-16}
Nafar~Sefiddashti,~M.~H.; Edwards,~B.~J.; Khomami,~B. \emph{Journal of
  Rheology} \textbf{2016}, \emph{60}, 1227--1244\relax
\mciteBstWouldAddEndPuncttrue
\mciteSetBstMidEndSepPunct{\mcitedefaultmidpunct}
{\mcitedefaultendpunct}{\mcitedefaultseppunct}\relax
\EndOfBibitem
\bibitem[Nafar~Sefiddashti \latin{et~al.}(2019)Nafar~Sefiddashti, Edwards, and
  Khomami]{sek-19}
Nafar~Sefiddashti,~M.~H.; Edwards,~B.~J.; Khomami,~B. \emph{Macromolecules}
  \textbf{2019}, \emph{52}, 8124--8143\relax
\mciteBstWouldAddEndPuncttrue
\mciteSetBstMidEndSepPunct{\mcitedefaultmidpunct}
{\mcitedefaultendpunct}{\mcitedefaultseppunct}\relax
\EndOfBibitem
\bibitem[Likhtman and McLeish(2002)Likhtman, and McLeish]{lm-02}
Likhtman,~A.~E.; McLeish,~T. C.~B. \emph{Macromolecules} \textbf{2002},
  \emph{35}, 6332--6343\relax
\mciteBstWouldAddEndPuncttrue
\mciteSetBstMidEndSepPunct{\mcitedefaultmidpunct}
{\mcitedefaultendpunct}{\mcitedefaultseppunct}\relax
\EndOfBibitem
\bibitem[Masubuchi \latin{et~al.}(2003)Masubuchi, Ianniruberto, Greco, and
  Marrucci]{maigm-03}
Masubuchi,~Y.; Ianniruberto,~G.; Greco,~F.; Marrucci,~G. \emph{The Journal of
  Chemical Physics} \textbf{2003}, \emph{119}, 6925--6930\relax
\mciteBstWouldAddEndPuncttrue
\mciteSetBstMidEndSepPunct{\mcitedefaultmidpunct}
{\mcitedefaultendpunct}{\mcitedefaultseppunct}\relax
\EndOfBibitem
\bibitem[Tzoumanekas and Theodorou(2006)Tzoumanekas, and Theodorou]{tt-06}
Tzoumanekas,~C.; Theodorou,~D.~N. \emph{Macromolecules} \textbf{2006},
  \emph{39}, 4592--4604\relax
\mciteBstWouldAddEndPuncttrue
\mciteSetBstMidEndSepPunct{\mcitedefaultmidpunct}
{\mcitedefaultendpunct}{\mcitedefaultseppunct}\relax
\EndOfBibitem
\bibitem[Foteinopoulou \latin{et~al.}(2006)Foteinopoulou, Karayiannis,
  Mavrantzas, and Kr{\"o}ger]{fkmk-06}
Foteinopoulou,~K.; Karayiannis,~N.~C.; Mavrantzas,~V.~G.; Kr{\"o}ger,~M.
  \emph{Macromolecules} \textbf{2006}, \emph{39}, 4207--4216\relax
\mciteBstWouldAddEndPuncttrue
\mciteSetBstMidEndSepPunct{\mcitedefaultmidpunct}
{\mcitedefaultendpunct}{\mcitedefaultseppunct}\relax
\EndOfBibitem
\bibitem[Baig \latin{et~al.}(2010)Baig, Mavrantzas, and Kr{\"o}ger]{bmk-10}
Baig,~C.; Mavrantzas,~V.~G.; Kr{\"o}ger,~M. \emph{Macromolecules}
  \textbf{2010}, \emph{43}, 6886--6902\relax
\mciteBstWouldAddEndPuncttrue
\mciteSetBstMidEndSepPunct{\mcitedefaultmidpunct}
{\mcitedefaultendpunct}{\mcitedefaultseppunct}\relax
\EndOfBibitem
\bibitem[Giesekus(1982)]{giesekus-82}
Giesekus,~H. \emph{Journal of Non-Newtonian Fluid Mechanics} \textbf{1982},
  \emph{11}, 69--109\relax
\mciteBstWouldAddEndPuncttrue
\mciteSetBstMidEndSepPunct{\mcitedefaultmidpunct}
{\mcitedefaultendpunct}{\mcitedefaultseppunct}\relax
\EndOfBibitem
\bibitem[Maklad and Poole(2021)Maklad, and Poole]{mp-21}
Maklad,~O.; Poole,~R.~J. \emph{Journal of Non-Newtonian Fluid Mechanics}
  \textbf{2021}, 104522\relax
\mciteBstWouldAddEndPuncttrue
\mciteSetBstMidEndSepPunct{\mcitedefaultmidpunct}
{\mcitedefaultendpunct}{\mcitedefaultseppunct}\relax
\EndOfBibitem
\bibitem[Doi and Edwards(1988)Doi, and Edwards]{DoiEdwards_Book}
Doi,~M.; Edwards,~S.~F. \emph{The Theory of Polymer Dynamics}; Oxford
  University Press, 1988\relax
\mciteBstWouldAddEndPuncttrue
\mciteSetBstMidEndSepPunct{\mcitedefaultmidpunct}
{\mcitedefaultendpunct}{\mcitedefaultseppunct}\relax
\EndOfBibitem
\bibitem[Thompson \latin{et~al.}(2022)Thompson, Aktulga, Berger, Bolintineanu,
  Brown, Crozier, in't Veld, Kohlmeyer, Moore, Nguyen, \latin{et~al.}
  others]{thompson2022lammps}
Thompson,~A.~P.; Aktulga,~H.~M.; Berger,~R.; Bolintineanu,~D.~S.; Brown,~W.~M.;
  Crozier,~P.~S.; in't Veld,~P.~J.; Kohlmeyer,~A.; Moore,~S.~G.; Nguyen,~T.~D.,
  \latin{et~al.}  \emph{Computer Physics Communications} \textbf{2022},
  \emph{271}, 108171\relax
\mciteBstWouldAddEndPuncttrue
\mciteSetBstMidEndSepPunct{\mcitedefaultmidpunct}
{\mcitedefaultendpunct}{\mcitedefaultseppunct}\relax
\EndOfBibitem
\bibitem[Dobson(2014)]{d-14}
Dobson,~M. \emph{The Journal of Chemical Physics} \textbf{2014}, \emph{141},
  184103\relax
\mciteBstWouldAddEndPuncttrue
\mciteSetBstMidEndSepPunct{\mcitedefaultmidpunct}
{\mcitedefaultendpunct}{\mcitedefaultseppunct}\relax
\EndOfBibitem
\bibitem[Nicholson and Rutledge(2016)Nicholson, and Rutledge]{nr-16}
Nicholson,~D.~A.; Rutledge,~G.~C. \emph{The Journal of Chemical Physics}
  \textbf{2016}, \emph{145}, 244903\relax
\mciteBstWouldAddEndPuncttrue
\mciteSetBstMidEndSepPunct{\mcitedefaultmidpunct}
{\mcitedefaultendpunct}{\mcitedefaultseppunct}\relax
\EndOfBibitem
\bibitem[O'Connor \latin{et~al.}(2019)O'Connor, Hopkins, and Robbins]{ohr-19}
O'Connor,~T.~C.; Hopkins,~A.; Robbins,~M.~O. \emph{Macromolecules}
  \textbf{2019}, \emph{52}, 8540--8550\relax
\mciteBstWouldAddEndPuncttrue
\mciteSetBstMidEndSepPunct{\mcitedefaultmidpunct}
{\mcitedefaultendpunct}{\mcitedefaultseppunct}\relax
\EndOfBibitem
\bibitem[Kr{\"o}ger(2005)]{kroger-05}
Kr{\"o}ger,~M. \emph{Computer Physics Communications} \textbf{2005},
  \emph{168}, 209--232\relax
\mciteBstWouldAddEndPuncttrue
\mciteSetBstMidEndSepPunct{\mcitedefaultmidpunct}
{\mcitedefaultendpunct}{\mcitedefaultseppunct}\relax
\EndOfBibitem
\bibitem[Kröger \latin{et~al.}(2023)Kröger, Dietz, Hoy, and Luap]{kdhl-23}
Kröger,~M.; Dietz,~J.~D.; Hoy,~R.~S.; Luap,~C. \emph{Computer Physics
  Communications} \textbf{2023}, \emph{283}, 108567\relax
\mciteBstWouldAddEndPuncttrue
\mciteSetBstMidEndSepPunct{\mcitedefaultmidpunct}
{\mcitedefaultendpunct}{\mcitedefaultseppunct}\relax
\EndOfBibitem
\bibitem[Siepmann \latin{et~al.}(1993)Siepmann, Karaborni, and Smit]{sks-93}
Siepmann,~J.~I.; Karaborni,~S.; Smit,~B. \emph{Nature} \textbf{1993},
  \emph{365}, 330--332\relax
\mciteBstWouldAddEndPuncttrue
\mciteSetBstMidEndSepPunct{\mcitedefaultmidpunct}
{\mcitedefaultendpunct}{\mcitedefaultseppunct}\relax
\EndOfBibitem
\bibitem[Ianniruberto and Marrucci(2014)Ianniruberto, and Marrucci]{im-14}
Ianniruberto,~G.; Marrucci,~G. \emph{Journal of Rheology} \textbf{2014},
  \emph{58}, 89--102\relax
\mciteBstWouldAddEndPuncttrue
\mciteSetBstMidEndSepPunct{\mcitedefaultmidpunct}
{\mcitedefaultendpunct}{\mcitedefaultseppunct}\relax
\EndOfBibitem
\bibitem[Ianniruberto and Marrucci(2014)Ianniruberto, and Marrucci]{im-14b}
Ianniruberto,~G.; Marrucci,~G. \emph{Journal of Rheology} \textbf{2014},
  \emph{58}, 1083--1083\relax
\mciteBstWouldAddEndPuncttrue
\mciteSetBstMidEndSepPunct{\mcitedefaultmidpunct}
{\mcitedefaultendpunct}{\mcitedefaultseppunct}\relax
\EndOfBibitem
\bibitem[Ianniruberto(2015)]{ianniruberto-15}
Ianniruberto,~G. \emph{Journal of Rheology} \textbf{2015}, \emph{59},
  211--235\relax
\mciteBstWouldAddEndPuncttrue
\mciteSetBstMidEndSepPunct{\mcitedefaultmidpunct}
{\mcitedefaultendpunct}{\mcitedefaultseppunct}\relax
\EndOfBibitem
\bibitem[Hawke \latin{et~al.}(2015)Hawke, Huang, Hassager, and Read]{hhhr-15}
Hawke,~L. G.~D.; Huang,~Q.; Hassager,~O.; Read,~D.~J. \emph{Journal of
  Rheology} \textbf{2015}, \emph{59}, 995--1017\relax
\mciteBstWouldAddEndPuncttrue
\mciteSetBstMidEndSepPunct{\mcitedefaultmidpunct}
{\mcitedefaultendpunct}{\mcitedefaultseppunct}\relax
\EndOfBibitem
\bibitem[Nafar~Sefiddashti \latin{et~al.}(2019)Nafar~Sefiddashti, Edwards, and
  Khomami]{nek-19a}
Nafar~Sefiddashti,~M.~H.; Edwards,~B.~J.; Khomami,~B. \emph{Polymers}
  \textbf{2019}, \emph{11}, 476\relax
\mciteBstWouldAddEndPuncttrue
\mciteSetBstMidEndSepPunct{\mcitedefaultmidpunct}
{\mcitedefaultendpunct}{\mcitedefaultseppunct}\relax
\EndOfBibitem
\end{mcitethebibliography}
\newpage


\end{document}


\singlespacing
\newpage
\section{Table of quantities}
\begin{table*}[htb!]
    \begin{xltabular}{\textwidth}{ccX}\hline\hline
    Symbol & Formula & Definition \\\hline
    $N_{\textrm{mon}}$&&  Number of united atoms per chain in the polyethylene melts or the number of bead per chain in the Kremer-Grest melts.\\[18truept]
    $\ell_b$ & & Length of an individual bond. \\[8truept]
    $b_K$ && Kuhn length of a polymer. \\[8truept]
     $L$ && Primitive-path length of a polymer, as obtained from the Z1 code. \\[8truept]
     $R_\textrm{max}$&& Maximum extension of a molecule with the bonds in the minimum energy configuration.\\[8truept]
     $\ens{R^2}$ && Mean square end-to-end length of a molecule. \\ [8truept]
          $Z_k$ &&   Number of topological entanglements, determined by counting the number of kinks in the Z1 code chain shrinking algorithms.\cite{sk-07} \\[18truept]
           $Z_{k,eq}$ && The equilibrium number of topological entanglements. \\[5truept]
          $Z$ or $Z_{\textrm{rheol}}$&
          $\begin{cases}\dfrac{\ens{L}^2}{\ens{R^2}}\\[15truept] \dfrac{Z_{k,eq}+1}{2} \end{cases}$
          & \parbox{10.0truecm}{\change{The} \delete{R}\change{r}heological number of entanglements.\\[20truept]
          Approximate relation obtained from Z1 code.} \\[35truept]
           $N_K$ & $\dfrac{R_{\textrm{max}}}{b_K}$ &  Number of Kuhn segments in the chain for Kremer-Grest chains.\\[15truept]
            $N_K$ & $\dfrac{R_{\textrm{max}}^2}{\ens{R^2}}$ &  Number of Kuhn segments in the chain for polyethylene chains.\\[15truept]
           $N_{eK}$ & $\dfrac{N_K}{Z_\textrm{rheol}}$ & Number of Kuhn segments in a \textit{rheological} tube segment.\\[15truept]
           $N_e^{\textrm{mon}}$ & $\dfrac{N_\textrm{mon}}{Z_{\textrm{rheol}}}$ & Number of monomers in a rheological tube segment. 
           \\[15truept]
           $C_{\infty}$ & $\dfrac{\left<R^2\right>}{\ell_b^2(N_{\textrm{mon}}-1)}$& The characteristic ratio. \\[15truept]
           $C_{\infty}$ & $\dfrac{b_K}{\ell_b}$& The characteristic ratio for Kremer-Grests melts\\[15truept]
           $\lambda$ & $\dfrac{\ens{L}}{L_{eq}}$ & The (dimensionless) primitive path stretch.\\[15truept]
         \hline\hline
         \end{xltabular}
         \caption{Table of symbols and quantities.}
    \label{tab:quantities}
         \end{table*}
         \begin{table*}[t!]
          \begin{xltabular}{\textwidth}{ccX}\hline\hline
    Symbol & Formula & Definition \\\hline
            $\lambda_{\textrm{max}}$ & $\sqrt{N_{eK}}$& The maximum stretch. This quantity is independent of the current number of entanglements.\cite{dolata2023} \\[8truept]
            $\tens{A}$
          && The conformation tensor defined as in \citet{dolata2023} to be proportional to the stress and consistent with the stress-optical rule. \\[8truept]
         $\nu$ 
         &$\dfrac{Z}{Z_{eq}}$& The entanglement ratio 
         \\ $h(\lambda)$ 
         && The spring potential of a tube segment, given by the Cohen approximation to the inverse Langevin function.\cite{cohen-91,sbm-09} \\[8truept]
         $\beta$ 
         && The CCR (convective constraint release) parameter, roughly inversely related to the number of retraction events necessary to remove an entanglement. \\[8truept]
         $\tau_e$ 
         && The Rouse relaxation time for a chain the size of an entanglement strand. Computed using correlations from
        \citet{ekfhs-20} for the Kremer-Grest melts.\\[8truept]
         $\tau_R$ 
         && The Rouse time, computed using 
         $\tau_R=\tau_e Z_{\textrm{rheol}}^2$ 
         for the Kramer-Grest melts. The united-atom polyethylene melts provide values of the Rouse time computed from the center-of-mass diffusivity using tube theory, and by identifying cross-over regimes in the segmental mean squared displacement.\cite{sek-15,sek-16,sek-19}  We find that the \change{tube theory} \delete{theoretical} values provide better fits for the polymer rheology, and use these values throughout. \\[8truept]
         $\tau_d$ 
         &Eq.~\ref{eq:tau-d-eq}& The reptation (disengagement) time. 
         computed from the Rouse time using the Likhtman\delete{-McLeish}\cite{lm-02} scaling relation. \\
         \hline\hline
    \end{xltabular}
    \end{table*}

\newpage
\section{Parameter Determination}

The number of tube segments $Z_\textrm{rheol}$ in the rheological tube  sets the plateau modulus according to
\begin{equation}\label{eq:plateau-modulus}
	G_0 = Z_\textrm{rheol}n\kB T,
\end{equation}
where $n$ is the number density of polymer chains, $\kB$ is Boltzmann constant, and $T$ is the temperature. The two definitions of entanglements are related by 
\begin{subequations}
\begin{align}
		\zeta_Z & = \frac{Z_{k,eq}}{Z_\textrm{rheol}} \label{eq:zeta}\\
			&  \simeq \frac{2}{1 + 1/Z_{k,eq}}\\ &\simeq 2,\label{eq:zetadef}
\end{align}
\end{subequations}
which arises from comparing the topological number of kinks $Z_{k,eq}$ with the rheological value $Z_{\textrm{rheol}}$.
\cite{maigm-03,tt-06,fkmk-06,bmk-10}   The equilibrium reptation time $\tau_{d,eq}$ can be computed from the rheological tube segments as expressed in Eq.~\eqref{eq:zeta} through the Likhtman relation \cite{lm-02}
\begin{equation}\label{eq:tau-d-eq}
	\frac{\tau_{d,eq}}{3Z_\textrm{rheol}} = \left(1 - \frac{3.38}{Z_\textrm{rheol}^{1/2}} + \frac{4.17}{Z_\textrm{rheol}} - \frac{1.55}{Z_\textrm{rheol}^{3/2}} \right)\tau_R.
\end{equation}
The Giesekus parameter\cite{giesekus-82} 
\begin{equation}\label{eq:alpha}
	\alpha \equiv -\lim_{\dot\gamma\to0}2\frac{\Psi_2}{\Psi_1}
\end{equation} 
is defined by the ratio of the second and first normal stress coefficients in the limit of vanishing shear rate.  This leaves the CCR parameter $\beta$ as the only free parameter, since all others ($\alpha, N_{eK}, Z_{k,eq}, \tau_R$) can be measured or computed.

Our prior work demonstrates that the best fit value of $\beta$ will be insensitive to $\alpha$ over the range $0.2<\alpha<0.5$ (\textit{c.f.} figure 6 of \citet{dolata2023}). This range is consistent with experimentally determined values of $\alpha$\cite{mp-21}, and is also consistent with the \change{values of $4/7$ and $2/7$} \delete{limits of $2/7$ and $4/7$} predicted from tube theory \change{with and without the independent alignment approximation (\textit{c.f.} Eq. 7.205 of \citet{DoiEdwards_Book})} . As such, any physically reasonable value of $\alpha$ will yield approximately the same best-fit value of $\beta$. In this work, we use $\alpha=0.5$ for all melts.  We inferred this value from \change{\citet{sek-19}} for the longest united-atom polyethylene (UA-PE) melts using \change{\eqref{eq:alpha},}
\eqdel{\begin{equation*}}
    \edelete{\lim_{\dot\gamma\to 0} \frac{N_2}{N_1} = -\frac{\alpha}{2},}
\eqdel{\end{equation*}}
and assume that it holds regardless of molecular weight.  
We use value $\alpha=0.5$ for the semi-flexible (FG-KG) and  flexible (F-KG) Kremer-Grest melts because direct measurement of $\alpha$ required unfeasibly long computation time. 
\delete{In prior work, we showed that the predictions of entanglement reduction are roughly unchanged  for  $0.2<\alpha<0.5$ \cite{dolata2023}; this range is consistent with measured second normal-stress differences in experiments.}

\section{Simulation methods}
All simulations of  Kremer-Grest bead-spring models were performed \change{using} \delete{on} LAMMPS \cite{thompson2022lammps}. All melts simulated were at \change{a number} density of $\delete{\rho}\change{n}=0.85/a^3$. Melts with $N=500$ monomers per chain had $M=368$ chains, while melts with $N=250$ monomers per chain had $M=736$ chains in total. \change{Semiflexible chains of lengths $N= 250$ and $N=500$; and flexible melts with $N=500$ were simulated.}

Monomers interact via a purely repulsive truncated Lennard-Jones (LJ) pair potential
 \begin{equation}
    U_{LJ} = \begin{cases}4\epsilon\left[\left(\dfrac{a}{r}\right)^{12}-\left(\dfrac{a}{r}\right)^6+\dfrac{1}{4}\right]&\qquad r < r_c = 2^{1/6}a\\
    0 &\qquad r>r_c
    \end{cases}
 \end{equation}
 Relevant quantities measured in simulations are given in LJ units: $m$ is the monomer mass, $a$ is the monomer diameter, $\epsilon$ is the interaction energy, and time is given in units of $\tau=\sqrt{ma^2/\epsilon}$. The covalent bonds between beads in the same chain are modeled with the FENE potential
 \begin{equation}
    U_{\textrm{FENE}} = -0.5KR_0^2\ln\left[1-(r/R_0)^2\right],
 \end{equation}
 where $K = 30\epsilon/a^2$ and $R_0 = 1.5a$. For the semi-flexible chains, an additional angle bending potential is used, $U_{\textrm{bend}} = k_{\textrm{bend}}(1-\cos{\theta})$, where $k_{\textrm{bend}} = 1.5\epsilon$ and $\theta$ is the angle between successive bonds.

 The KG simulations were carried out using the SLLOD  algorithm as implemented in the Large-scale Atomic/Molecular Massively Parallel Simulator (LAMMPS) and temperature was kept at a constant $T/\epsilon = 1$. SLLOD works by imposing a uniform shear velocity profile across the simulation box, and therefore is not able to accurately describe inhomogeneous behavior such as shear banding. During shear, temperature was controlled by a Nos\'e-Hoover thermostat with a damping time of 10$\tau$. During relaxation after cessation of shear, temperature was controlled by a Langevin thermostat that subtracts the remaining velocity profile from the shear stage during the early stages of relaxation, before adjusting the temperature. Different protocols during cessation (Nos\'e-Hoover thermostatting, not removing the velocity profile and thermostatting only velocities perpendicular to shear plane) were tested and found to not significantly alter the results for the rates studied.

 Extensional flow simulations utilize generalized Kraynik-Reinelt boundary conditions to avoid the problem of the simulation box becoming too narrow at large strains\cite{d-14,nr-16,ohr-19}. The temperature control and damping time were the same as those used in the shear simulation.

 `Kinks' were calculated using the Z1 code\cite{kroger-05}, which identifies topological constraints between polymers and outputs the primitive path. This method differs from the traditional PPA because it is not based on molecular dynamics, but on a series of topological moves that reduce the chains to their topological constraints. A more recent version of the code \cite{kdhl-23} is available, but was not used in these calculations. At equilibrium, the number of kinks $Z_k$ is roughly proportional to the number of rheological entanglements, $Z_k\simeq 2Z_{\textrm{rheo}}.$ 

Our methods were equivalent to the methods employed in the UA-PE simulations.\cite{sek-15,sek-16,sek-19} The UA-PE simulations used the SKS (Siepmann-Karaborni-Smit) potential\cite{sks-93} and were performed in LAMMPS using the SLLOD algorithm and temperature was controlled at $T=450 \,\textrm{K}$ using a Nos\'e-Hoover thermostat. The number of entanglements was computed using the Z1 code.

\section{Comparison with the Ianniruberto-Marruci Model}
\change{\subsection{Cessation of a Steady Shear Flow}}
We compare the results of \delete{the} our model and the Ianniruberto-Marruci (IM) model following the cessation of a \change{steady} shear flow.  Ianniruberto and Marrucci \cite{im-14,im-14b} calculated the stretch dynamics according to
\begin{equation}
    \frac{\p \lambda}{\p t} = -\lambda\tens{S}:\bnabla\vec{v} - \frac{\lambda-\nu^{1/2}}{\tau_R}.
\end{equation}
where the orientation tensor $\tens{S}$ was calculated using the Doi-Edwards model, with the reptation rate replaced by its modification due to CCR, 
\begin{equation}
\frac{1}{\tau_d(\beta)} = \frac{1}{\tau_{d,eq}} + \beta\,\mathbf{\nabla v:S}\change{,}\delete{.}
\end{equation}
\change{and with the affine deformation computed using a Seth tensor}. We will compute the entanglement dynamics in the IM model using both their original expression where entanglements recover on the reptation time and a modified expression consistent with the Rouse relaxation of our model:
\begin{subequations}
    \begin{align}
   &\textrm{IM Original:} &\frac{\p\nu}{\p t} &= -\beta\nu\left(\mathbf{\nabla v:S}-\frac{d\ln\lambda}{dt}\right) - \frac{\nu-1}{\tau_d},
\label{eq:IM-nu}\\
  & \textrm{IM Modified:}& \frac{\p\nu}{\p t}& = -\beta\nu\left(\mathbf{\nabla v:S}-\frac{d\ln\lambda}{dt}\right) - \frac{\ln\nu}{\tau_R},
    \end{align}
\end{subequations}

\begin{figure}
    \centering\includegraphics{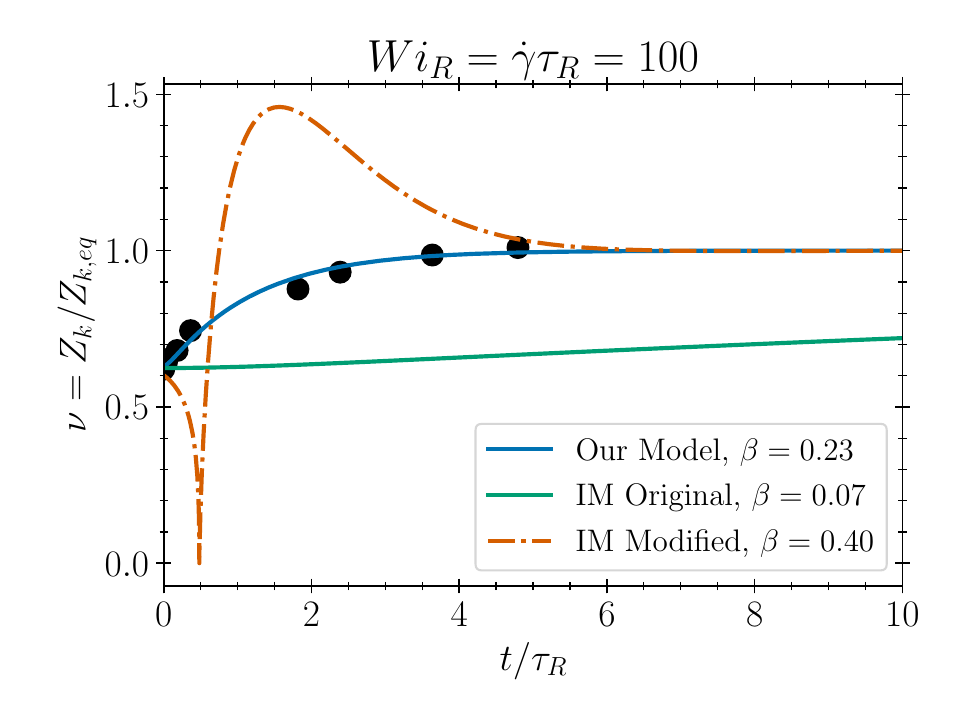}
    \caption{Comparison between our model and the IM model for re-entanglement of the SF-KG \delete{500} melt \change{with 500 monomers} following cessation of a steady shear. Lines represent theoretical calculations and symbols represent simulation results.}\label{fig:IMComparison}
\end{figure}

In figure \ref{fig:IMComparison}, we compare simulated results for re-entanglement of the SF-KG 500 melt following cessation of a flow at ${Wi}_R=100$; in the IM-model, we choose $\beta$ to give best fit \change{of $\nu$ in} \delete{at the} steady state flow.  We find that our model describes the re-entanglement of the melt accurately, while the IM model with re-entanglement on the reptation time re-entangles too slowly.  The IM model with re-entanglement on the Rouse time is unstable; the number of entanglements initially drops to zero, grows to a value above equilibrium, and then decays towards equilibrium on the Rouse time.  This behavior arises from the form of the\delete{se} modified IM equations during retraction after the cessation of the flow:
\begin{subequations}
    \begin{align}
        \frac{\p\lambda}{\p t} & = -\frac{\lambda - \nu^{1/2}}{\tau_R}, \label{eq:lambda-relaxation} \\
        \frac{\p\nu}{\p t} & = \frac{\beta\nu}{\lambda}\frac{\p\lambda}{\p t} -\frac{\ln\nu}{\tau_R}.   
    \end{align}
\end{subequations}
Here, the initially stretched primitive path will contract, removing entanglements at the chain ends\change{. This disentanglement increases $\p\lambda/\p t$ in \eqref{eq:lambda-relaxation}}, causing further disentanglement.  If $\beta$ is sufficiently large, this continues until the number of entanglements and $\p\lambda/\p t$ approaches zero, giving rise to a strong thermodynamic driving force that increases the number of entanglements.  Further refinements of the IM model that introduced a multimode version of the stretch relaxation equation\cite{ianniruberto-15} may resolve the instability observed during relaxation.

Essentially, the instability is driven by the assumption that the stretch relaxes to a locally-equilibrated tube of length $\nu^{1/2}$.  This assumption, while reasonable on its face, is contradicted by the observation that the relaxation of stretch under the addition and removal of entanglements is asymmetric due to the fast relaxation of Kuhn segments following entanglement removal.\cite{hhhr-15} This asymmetry implies that the addition and removal of entanglements is an inherently dissipative process, and so $\lambda=\nu^{1/2}$ cannot be regarded as a local equilibrium for the stretch.  Furthermore, the couplings between $\nu$ and $\lambda$ in Eq.~\eqref{eq:lambda-relaxation} are  forbidden by Onsager-Casimir reciprocity.\cite{dolata2023} The considerations highlight the importance of incorporating thermodynamic constraints in rheological constitutive equations.

\begin{figure}
\centering
\includegraphics[width=0.55\textwidth]{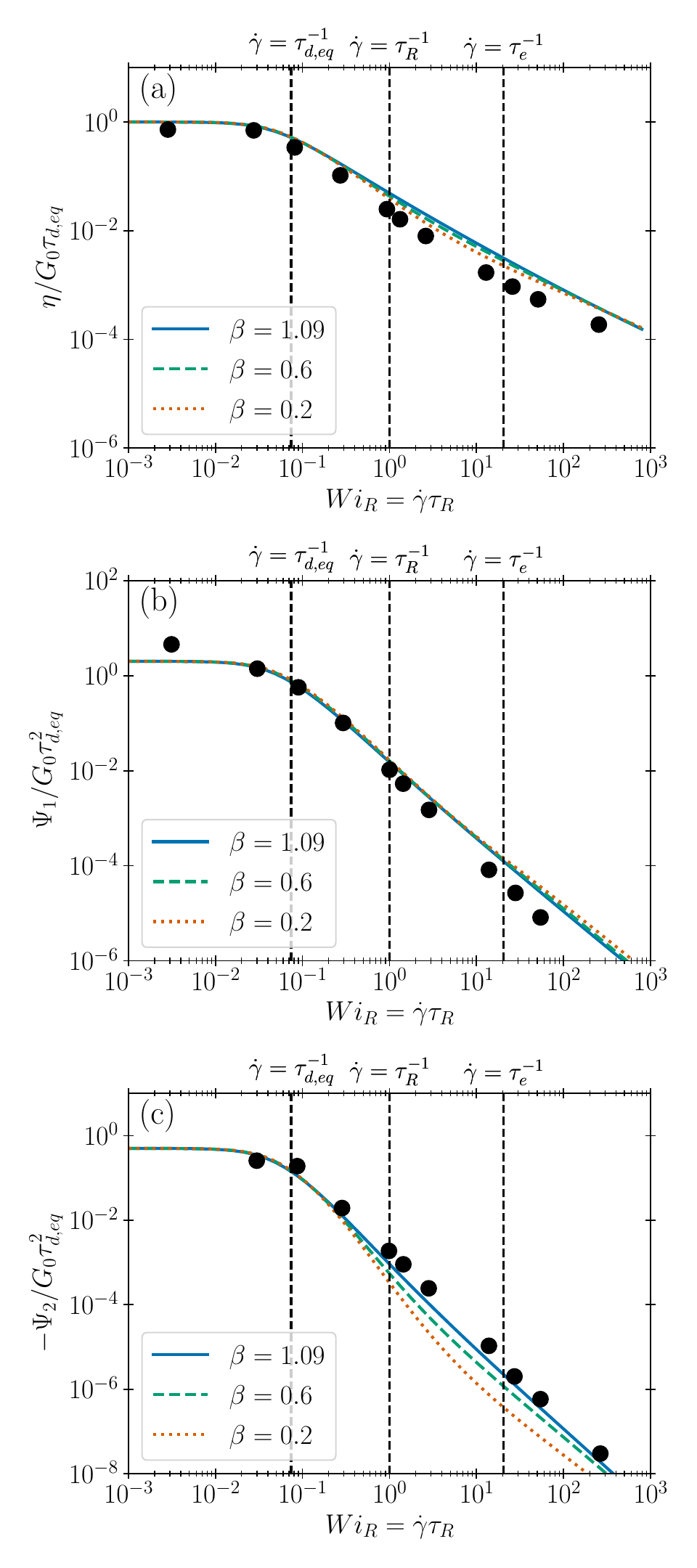}
	\caption{(a) Steady-shear viscosity $\eta$, (b) first normal stress coefficient $\Psi_1$, and (c) second normal stress coefficient $\Psi_2$ as a function of $\Wi_R$.  Filled circles are UA-PE with $Z_{k,eq}=24.8$,\cite{sek-19} solid lines represent our model.}\label{fig:steady-state}
\end{figure}

\subsection{\change{Steady-State} Rheological Predictions}
Fig.~\ref{fig:steady-state} compares  predictions of our model \change{for the shear viscosity} with the shear viscosity of the longest UA-PE chains\cite{sek-19},  for  $\beta=0.2,0.6,1.0$, where $\beta=1.0$ provided the best fit to shear-induced disentanglement.  We see good agreement \change{for all $\beta$} between the model predictions (solid lines)  \delete{for all $\beta$} for the shear viscosity and the simulations (filled circles) over the entire range of ${\Wi}_R$, with discrepancies increasing as the shear rate approaches the inverse entanglement time $\tau_e^{-1}$.  On these timescales, simulations of UA-PE have shown that polymer molecules undergo cycles of retraction and extension\cite{nek-19a} that are not described by our model, which may contribute to an over-prediction of the viscosity.  The viscosity and first normal-stress coefficient are fairly insensitive to the value of the CCR parameter $\beta$, which suggests that the shear viscosity is a weak discriminator for $\beta$.  Interestingly, the best fit of $\beta$ obtained from the UA-PE method via Method A also provides the best fit for the thinning of the second normal stress coefficient.  This suggests that $\Psi_2$ could be a better discriminant for $\beta$ than the other rheometric coefficients.

\section{Disclaimer}
Certain commercial or open-source software are identified in this paper in order to specify the methodology adequately. Such identification is not intended to imply recommendation or endorsement of any product or service by NIST, nor is it intended to imply that the software identified are necessarily the best available for the purpose.

\bibliography{GENERIC-ref.bib}